\documentclass[aps,pra,twocolumn,floatfix,showpacs,preprintnumbers,amsmath,amssymb,10pt]{revtex4-1}

\DeclareMathAlphabet{\mathsfit}{\encodingdefault}{\sfdefault}{m}{sl}
\SetMathAlphabet{\mathsfit}{bold}{\encodingdefault}{\sfdefault}{bx}{sl}

\usepackage[dvips]{graphics}
\usepackage{mathptmx}
\usepackage{psfrag}
\usepackage{graphicx}
\usepackage{bm}
\usepackage{amsmath}
\usepackage{trfsigns}
\usepackage{amssymb}
\usepackage[usenames,dvipsnames]{color}
\usepackage{dashrule}
\usepackage{textgreek}
\usepackage[english]{babel}
\usepackage{contour}
\usepackage{upgreek,bm}
\usepackage{color}
\definecolor{dred}{rgb}{.6,.0,0.}
\definecolor{dblue}{rgb}{.0,.0,0.6}

\renewcommand{\vec}[1]{\mathbf{#1}}
\newcommand{\tens}[1]{\mbox{\textsf{\textbf{#1}}}}
\newcommand{\Greektens}[1]{\contour[3]{black}{#1}}
\newcommand{\veczero}{\mathbf{0}}
\newcommand{\tenszero}{\mbox{\textsf{\textbf{0}}}}
\newcommand{\sprod}{\!\cdot\!}
\newcommand{\tprod}{}
\newcommand{\vprod}{\!\times\!}

\newcommand{\trans}{{\operatorname{T}}}

\newcommand{\dif}{\mathrm{d}}
\newcommand{\mi}{\textrm{i}} 
\newcommand{\me}{\mathrm{e}}

\begin{document}

\title{Casimir--Polder Shift and Decay Rate in the Presence of Nonreciprocal Media}

\author{Sebastian Fuchs$^1$}
\author{J. A. Crosse$^2$}
\author{Stefan Yoshi Buhmann$^{1,3}$}
\affiliation{$^1$ Physikalisches Institut, Albert-Ludwigs-Universit\"at Freiburg, Hermann-Herder-Stra{\ss}e 3, 79104 Freiburg, Germany\\
$^2$ Department of Electrical and Computer Engineering, National University of Singapore, 4 Engineering Drive 3, 117583, Singapore\\
$^3$ Freiburg Institute for Advanced Studies, Albert-Ludwigs-Universit\"at Freiburg, Albertstra{\ss}e 19, 79104 Freiburg, Germany}

\date{\today}

\begin{abstract}
We calculate the Casimir--Polder frequency shift and decay rate for an atom in front of a nonreciprocal medium by using macroscopic quantum electrodynamics. The results are a generalization of the respective quantities for matter with broken time-reversal symmetry which does not fulfill the Lorentz reciprocity principle. As examples, we contrast the decay rates, the resonant and nonresonant frequency shifts of a perfectly conducting (reciprocal) mirror to those of a perfectly reflecting nonreciprocal mirror. We find different power laws for the distance dependence of all quantities in the retarded and nonretarded limits. As an example of a more realistic nonreciprocal medium, we investigate a topological insulator subject to a time-symmetry breaking perturbation.
\end{abstract}


\maketitle

\section{Introduction}
The Casimir--Polder force \cite{Casimir_Polder:1948}, like the van der Waals and the Casimir forces \cite{Casimir:1948}, is a dispersion force \cite{Buhmann_Book_1, Buhmann_Book_2}. This weak electromagnetic force is a result of noise currents composed of noise polarization and noise magnetization in matter, which are described by macroscopic electric and magnetic quantities. The noise currents act as a source for a quantized electromagnetic field, which can be expanded in terms of the classical electromagnetic Green's tensor for the Helmholtz equation \cite{Dung:1998, Buhmann:2004, Buhmann:2007, Scheel:2008, Skagerstam:2009, Chew_Book}. By computing the interaction of an atom with this field one can compute the effect of material bodies on the internal properties of the atom. The Casimir--Polder force is a result of the level shift of the atom induced by this field. In this theoretical framework, materials are described macroscopically by electromagnetic physical quantities and therefore this approach is known as macroscopic quantum electrodynamics (QED) \cite{Knoll_Book, Buhmann_Book_1, Buhmann_Book_2}.\\
Casimir--Polder potentials have been investigated for graphene \cite{Eberlein:2012}, metamaterials~\cite{Xu:2014, Henkel:2005} and Rydberg atoms near metallic surfaces \cite{Crosse:2010}. In the theory of macroscopic QED, Lorentz reciprocity principle stating the reversibility of optical paths, i.e. the symmetry with respect to an exchange of positions and orientations of sources and fields, holds for reciprocal material (Lorentz reciprocity being a particular case of the Onsager reciprocity from statistical physics \cite{Onsager:1931}). Thus these materials preserve the time-reversal symmetry. In order to study Casimir--Polder potentials for nonreciprocal media, which violate Lorentz reciprocity relation \cite{Lorentz:1896}, the theory of macroscopic QED was generalized to include cross-susceptibilities which mix the electric and magnetic fields \cite{Buhmann_nonreciprocal:2012}. Cross-susceptibilities arise, for instance, in topological insulators, but also in strictly reciprocal media such as chiral (meta-)materials \cite{Yannopapas:2006, Thiel:2009}. In this paper, we investigate the Casimir--Polder frequency shift and decay rate for a nonreciprocal medium.\\
Topological insulators \cite{Shen:2014,Bernevig:2006,Hasan:2010} are time-symmetric materials which are characterized by an insulating bulk and protected conducting surface states and have been observed in 3D in materials which exhibit strong enough spin-orbit coupling to induce band inversion \cite{Zhang:2009}. These materials can be used to realize axion media. To do this one needs to introduce a time-reversal symmetry breaking perturbation to the surface, either via ferromagnetic dopants \cite{Qi:2009, Chen:2010} or an external static magnetic field \cite{Maciejko:2010}. Such a perturbation opens a gap on the surface converting the surface conductor into a full insulator and leads to a non-trivial electromagnetic response --- in particular the electric, E, and magnetic induction, B, fields are able to mix \cite{Qi:2009}.\\
This magneto-electric effect can be described by adding an axion Lagrangian density term $\mathcal{L}_{\textrm{axion}} = \alpha/(4 \pi^2) \theta (\vec{r}, \omega) \vec{E} \sprod \vec{B}$ to the usual electro-magnetic Lagrangian density \cite{Wilczek:1987}. Here, $\alpha$ is the fine structure constant and $\theta(\mathbf{r},\omega)$ is the space- and time- dependent axion coupling. The axion coupling, $\theta$, vanishes in a trivial insulator but takes odd integer values of $\pi$ in a time-symmetry broken topological insulator, with the value and sign of the integer related to the strength and direction of the time-symmetry breaking perturbation. Physically, this describes a quantum Hall effect on the surface of the topological insulator \cite{Qi:2009}. The lowest Hall plateau, which is obtainable with an infinitesimally small perturbation, leads to an axion coupling of $\pm\pi$. Increasing the size of the perturbation will not change the axion coupling until the next Hall plateau is reached, where upon the axion coupling will increase to $\pm 3\pi$. Larger perturbations would result in even higher axion couplings as the relevant Hall plateaus are reached. It has been previously shown that the mixing of the electric and magnetic fields by the axion coupling has a significant effect on the Casimir force \cite{Grushin:2011} and, as we will show in Sec.~\ref{sec:Applications and Results}, it also modifies the Casimir--Polder shift.\\
An axion coupling does not only emerge in topological insulators, but also in metamaterials. Optical properties, e.g. reflective and transmissive properties, Fresnel formula, Brewster angle and the Goos--H\"{a}nchen effect of these materials have been studied theoretically in Refs.~\cite{Chang:2009, Zuo:2013}. As for layered topological insulators with a time-reversal symmetry breaking perturbation, potential applications are broad, for example a waveguide that induces polarization rotations due to the magneto-electric effect and mixes the electric and magnetic induction fields at the material's surface \cite{Crosse_2:2015}.\\
In our context, Casimir repulsion is of specific interest, e.g. the Casimir repulsion for magnetodielectric metamaterials predicted in Refs.~\cite{Rosa:2008, Rosa:2008_2, Zhao:2009}. Specifically, repulsive dispersion forces for a setup containing topological insulators are discussed in Ref.~\cite{Grushin:2011}, such as Casimir forces between three-dimensional topological insulators. Based on this approach, it is shown in Ref.~\cite{Chen:2011} that there is a critical band-gap where the Casimir force switches from attractive to repulsive. The Casimir--Polder interaction between an atom and a graphene surface with an applied magnetic field is studied in Ref.~\cite{Cysne:2014}. The authors observe plateau-like discontinuities of the Casimir--Polder interaction energy for specific values of the magnetic field and at low temperatures. This effect is traced back to the quantum Hall effect and is thus closely connected to our approach. We are going to apply the extended theory of macroscopic QED for nonreciprocal media to calculate frequency shifts and atomic decay rates of an atom in front of a topological insulator by directly using the electromagnetic properties derived in Ref.~\cite{Crosse:2015}.\\
This paper on the Casimir--Polder shift and decay rate in the presence of nonreciprocal media is organized as follows: The time-dependent electric field is calculated in the framework of macroscopic QED for nonreciprocal media in Sec.~\ref{sec:Electric Field}. This result is reached alternatively by a direct quantization of the noise current or by expressing noise polarization and magnetization through electromagnetic response functions and is needed for studying the internal atomic dynamics. This is described in Sec.~\ref{sec:Internal Dynamics} where the modified equations for the frequency shift and decay rate for nonreciprocal media are presented. In Sec.~\ref{sec:Applications and Results}, the results are applied to a perfectly reflecting nonreciprocal mirror and a topological insulator described by an axion coupling. In this context, we distinguish between a pure nonreciprocal topological insulator and material properties similar to $\textrm{Bi}_2\textrm{Se}_3$. Finally, we discuss the possibility of switching between attractive and repulsive Casimir--Polder force.

\section{The time-dependent electric field}
\label{sec:Electric Field}
A nonreciprocal medium violates time-reversal symmetry and, hence, the Lorentz reciprocity principle \cite{Lorentz:1896} for the Green's tensor does not hold:
\begin{equation}
\tens{G}^{\trans} \left( \vec{r} \: ', \vec{r}, \omega \right) \neq \tens{G} \left( \vec{r}, \vec{r} \: ', \omega \right).
\label{eq:Onsager's principle}
\end{equation}  
This necessitates new definitions for the real and imaginary parts of the Green's tensor $\tens{G}$
\begin{eqnarray}
\Re \left[ \tens{G} \left( \vec{r}, \vec{r} \: ' \right) \right] &=& \frac{1}{2} \left[ \tens{G} \left( \vec{r}, \vec{r} \: ' \right) + \tens{G}^{*\trans} \left( \vec{r} \: ', \vec{r} \right)\right]\\
\Im \left[ \tens{G} \left( \vec{r}, \vec{r} \: ' \right) \right] &=& \frac{1}{2 \mi} \left[ \tens{G} \left( \vec{r}, \vec{r} \: ' \right) - \tens{G}^{*\trans} \left( \vec{r} \: ', \vec{r} \right) \right].
\label{eq:Imaginary Part}
\end{eqnarray}
Thus the violation of Lorentz's principle calls for a modified mathematical description of macroscopic quantum electrodynamics (QED) for nonreciprocal media. Whereas the framework of macroscopic QED is described in Refs.~\cite{Buhmann_Book_1, Knoll_Book}, the modified approach for nonreciprocal media is outlined in Ref.~\cite{Buhmann_nonreciprocal:2012}. The internal dynamics of an atom with reciprocal media is discussed in Refs.~\cite{Buhmann_Book_2, Buhmann:2004}.\\
The general expression for the electric field reads
\begin{equation}
\hat{\vec{E}} \left( \vec{r} \right) = \int\limits^{\infty}_0{\dif \omega \: \left[ \hat{\vec{E}} \left( \vec{r}, \omega \right) + \hat{\vec{E}}{}^{\dagger} \left( \vec{r}, \omega \right) \right]}
\label{eq:Electric Field}
\end{equation}
with frequency components in Fourier space
\begin{align}
\begin{array}{lll}
&\hat{\vec{E}} \left( \vec{r}, \omega \right) &= \mi \mu_0 \omega \left[ \tens{G} \star \hat{\vec{j}}_{\textrm{N}} \right] \left( \vec{r}, \omega \right)\\[2mm]
& &= \mi \mu_0 \omega \displaystyle{\int{\dif ^3 r \: ' \: \tens{G} \left( \vec{r}, \vec{r} \: ', \omega \right) \cdot \hat{\vec{j}}_{\textrm{N}} \left( \vec{r} \: ', \omega \right)}},
\label{eq:Electric Field Components}
\end{array}
\end{align}
where $\star$ denotes a spatial convolution. The noise current density $\hat{\vec{j}}_{\textrm{N}}$ is governed by the quantum fluctuations occurring in the medium and has vanishing average $\langle \hat{\vec{j}}_{\textrm{N}} \rangle = \veczero$. $\hat{\vec{j}}_{\textrm{N}}$ can either be quantized directly, as is outlined in Sec.~\ref{sec:General Derivation}; or it can be represented by noise polarization $\hat{\vec{P}}_{\textrm{N}}$ and magnetization $\hat{\vec{M}}_{\textrm{N}}$ and the respective electric and magnetic fields are quantized separately yielding creation and annihilation operators for each field. We dedicate Sec.~\ref{sec:Concrete Derivation} to the second method using electric and magnetic response functions.\\
In order to obtain an expression for the time-dependent electric field \eqref{eq:Electric Field}, we have to find a solution for the time-dependent creation and annihilation operators first. This procedure is carried out both for a noise-current based schema and a polarization-magnetization founded method. The Hamiltonian $\hat H$ for the atom-field system is composed of the atomic part $\hat H_{\textrm{A}}$, the field part $\hat H_{\textrm{F}}$ and a contribution for the atom-field interaction $\hat H_{\textrm{AF}}$: $\hat H = \hat H_{\textrm{A}} + \hat H_{\textrm{F}} + \hat H_{\textrm{AF}}$. The atomic part $\hat{H}_{\textrm{A}}$
\begin{equation}
\hat{H}_{\textrm{A}}=\sum\limits_n{E_n \hat{A}_{nn}}
\label{eq:Atomic Hamiltonian}
\end{equation}
incorporates the eigenenergy $E_n$ for each atomic energy level and the atomic flip operator $\hat A_{mn} = |m \rangle \langle n|$. Resembling a harmonic oscillator, $\hat{H}_{\textrm{F}}$ comprises the integral over all the frequency-dependent number operators of the field-medium system and can be cast in the two aforementioned ways, cf. Secs.~\ref{sec:General Derivation} and \ref{sec:Concrete Derivation}. The interaction Hamiltonian $\hat{H}_{\textrm{AF}}$, which couples the atomic dipole to the electromagnetic field, reads
\begin{equation}
\hat H_{\textrm{AF}} =- \hat{\vec{d}} \sprod \hat{\vec{E}} \left( \vec{r}_A \right) =- \sum\limits_{m,n} {\hat{A}_{mn} \vec{d}_{mn} \sprod \hat{\vec{E}} \left( \vec{r}_A \right)}
\label{eq:Interaction Hamiltonian}
\end{equation}
and contains the electric-dipole operator $\hat{\vec{d}}=\sum_{m,n}{\vec{d}_{mn} \hat{A}_{mn}}$. Since $\hat{H}_{\textrm{A}}$ commutes with the field operators, only the commutation relations for the field Hamiltonian $\hat{H}_{\textrm{F}}$ and $\hat{H}_{\textrm{AF}}$ have to be studied to find the expression for the electric field \eqref{eq:Electric Field}. The field operators' equations of motion will be solved in the two different ways and inserted into Eq.~\eqref{eq:Electric Field Components}, thus giving a final expression for the electric field in the presence of an atom.
\subsection{General derivation for the noise current based schema in nonlocal media}
\label{sec:General Derivation}
In this first approach, the noise current is quantized directly by expressions for the field operators. Ohm's Law in frequency space
\begin{equation}
\hat{\vec{j}}_{\textrm{in}} \left( \vec{r}, \omega \right) = \left[ \tens{Q} \star \hat{\vec{E}} \right] \left( \vec{r}, \omega \right) + \hat{\vec{j}}_{\textrm{N}} \left( \vec{r}, \omega \right)
\end{equation}
describes the effect of the electric field $\hat{\vec{E}} \left( \vec{r}, \omega \right)$ on a linearly responding medium where $\tens{Q}$ is the conductivity matrix. Hence, the Helmholtz equation reads
\begin{multline}
\left[ \overrightarrow{\vec{\nabla}} \vprod \overrightarrow{\vec{\nabla}} \vprod - \frac{\omega^2}{c^2} \right] \tens{G} \left( \vec{r}, \vec{r} \: ', \omega \right) - \mi \mu_0 \omega \left[ \tens{Q} \star \tens{G} \right] \left( \vec{r}, \vec{r} \: ' \omega \right)\\
= \textrm{\Greektens{$\updelta$}} \left( \vec{r} - \vec{r} \: ' \right).
\end{multline}
This equation is formally solved by the Green's tensor $\tens{G}$ with $\tens{G} \rightarrow \tenszero$ for $|\vec{r} - \vec{r} \: '| \rightarrow \infty$.\\
We quantize the noise current density $\hat{\vec{j}}_{\textrm{N}}$ in Eq.~\eqref{eq:Electric Field Components} directly by writing it in terms of creation and annihilation operators $\hat{\vec{f}}{}^{\dagger}$ and $\hat{\vec{f}}$
\begin{equation}
\hat{\vec{j}}_{\textrm{N}} \left( \vec{r}, \omega \right) = \sqrt{\frac{\hbar \omega}{\pi}} \left[ \tens{R} \star \hat{\vec{f}} \right] \left( \vec{r}, \omega \right),
\label{eq:Noise current density}
\end{equation}
where $\tens{R}$ is related to the real part of the conductivity tensor $\tens{Q}$
\begin{equation}
\left[ \tens{R} \star \tens{R}^{*\trans} \right] \left( \vec{r}, \vec{r} \; ', \omega \right) = \Re \left[ \tens{Q} \left( \vec{r}, \vec{r} \; ', \omega \right) \right].
\label{eq:Identity Conductivity Matrix}
\end{equation}
The Heisenberg equation of motion for the annihilation operator $\hat{\vec{f}}$ and $\hat{\vec{f}}{}^{\dagger}$
\begin{equation}
\dot{\hat{\vec{f}}} \left( \vec{r}, \omega \right) = { \displaystyle{\frac{1}{\mi \hbar}}} \left[ \hat{\vec{f}} \left( \vec{r}, \omega \right), \hat H \right],
\label{eq:Heisenberg Equations of Motion}
\end{equation}
upon using the field Hamiltonian $\hat{H}_{\textrm{F}}$
\begin{equation}
\hat H_{\textrm{F}} = \int{\dif ^3 r \: \int\limits^{\infty}_0 {\dif \omega \: \hbar \omega \hat{\vec{f}}{}^{\dagger} \left( \vec{r}, \omega \right) \sprod \hat{\vec{f}} \left( \vec{r}, \omega \right) }}
\label{eq:Field Hamiltonian}
\end{equation}
and the interaction Hamiltonian $\hat{H}_{\textrm{AF}}$ \eqref{eq:Interaction Hamiltonian} by using Eq.~\eqref{eq:Electric Field Components}
\begin{multline}
\hat H_{\textrm{AF}} =-\sum\limits_{m,n} {\int\limits^{\infty}_0{\dif \omega \: \mi \mu_0 \omega \sqrt{\frac{\hbar \omega}{\pi}} \hat{A}_{mn} \vec{d}_{mn}}}\\
\left\{ \left[ \tens{G} \star \tens{R} \star \hat{\vec{f}} \right] \left( \vec{r}_A, \omega \right) - \left[ \tens{G}^* \star \tens{R}^* \star \hat{\vec{f}}{}^{\dagger} \right] \left( \vec{r}_A, \omega \right) \right\}
\label{eq:Interaction Hamiltonian1}
\end{multline}
gives the solution of the annihilation operator $\hat{\vec{f}}$
\begin{multline}
\hat{\vec{f}} \left( \vec{r}, \omega, t \right) = \me^{-\mi \omega \left( t-t_0 \right)} \hat{\vec{f}} \left( \vec{r}, \omega \right) + \frac{\mu_0 \omega}{\hbar} \sqrt{\frac{\hbar \omega}{\pi}} \sum\limits_{m,n}{\int\limits^{t}_{t_0}{\dif t'}}\\
{\me^{-\mi \omega \left( t-t' \right)}\left[ \tens{G} \star \tens{R} \right]^{*\trans} \left( \vec{r}_A, \vec{r}, \omega \right) \cdot \vec{d}_{mn} \hat A_{mn}}.
\label{eq:Equation of motion annihilation operator}
\end{multline}
Substituting the results into Eq.~\eqref{eq:Noise current density}, using Eq.~\eqref{eq:Identity Conductivity Matrix} and the expression 
\begin{equation}
\Im \left[ \tens{G} \left( \vec{r}, \vec{r} \: ', \omega \right) \right] = \mu_0 \omega \left[ \tens{G} \star \Re \left[ \tens{Q} \right] \star \tens{G}^{*\trans} \right] \left( \vec{r}, \vec{r} \: ', \omega \right)
\label{eq:Identity Imaginary Part Green's Tensor}
\end{equation}
leads to an expression for the electric field in nonreciprocal media
\begin{multline}
\hat{\vec{E}} \left( \vec{r}, \omega, t \right) = \me^{-\mi \omega \left( t-t_0 \right)} \hat{\vec{E}} \left( \vec{r}, \omega \right)\\
+ \mi \frac{\mu_0 \omega^2}{\pi} \sum\limits_{m,n}{\int\limits^t_{t_0}{\dif t' \: \me^{-\mi \omega \left( t-t' \right)} \Im \left[ \tens{G} \left( \vec{r}, \vec{r}_A, \omega \right) \right] \sprod \vec{d}_{mn} \hat A_{mn}}},
\label{eq:Results Components Electric Field1}
\end{multline}
which differs from the usual expression for reciprocal media only by the definition of the imaginary part of the Green's tensor \eqref{eq:Imaginary Part} \cite{Buhmann_Book_2}.

\subsection{Electric field in the polarization-magnetization based schema}
\label{sec:Concrete Derivation}
The components of the electric fields $\hat{\vec{E}} \left( \omega, t \right)$ can also be calculated in terms of electric and magnetic response functions, i.e. polarization and magnetization \cite{Buhmann_nonreciprocal:2012}. The constitutive relations for the electric displacement field $\hat{\vec{D}}$ and the magnetic induction field $\hat{\vec{B}}$ are given by \cite{Buhmann_nonreciprocal:2012}
\begin{alignat}{3}
&\hat{\vec{D}} &=& \epsilon_0 \textrm{\Greektens{$\upepsilon$}} \star \hat{\vec{E}} + \frac{1}{c} \textrm{\Greektens{$\upxi$}} \star \hat{\vec{H}} + \hat{\vec{P}}_{\textrm{N}} + \frac{1}{c} \textrm{\Greektens{$\upxi$}} \star \hat{\vec{M}}_{\textrm{N}}\\
&\hat{\vec{B}} &=& \frac{1}{c} \textrm{\Greektens{$\upzeta$}} \star \hat{\vec{E}} + \mu_0 \textrm{\Greektens{$\upmu$}} \star \hat{\vec{H}} + \mu_0 \textrm{\Greektens{$\upmu$}} \star \hat{\vec{M}}_{\textrm{N}},
\label{eq:Constitutive Relations}
\end{alignat}
where the tensor $\textrm{\Greektens{$\upepsilon$}}$ is the permittivity, $\textrm{\Greektens{$\upmu$}}$ the permeability and $\textrm{\Greektens{$\xi$}}$ and $\textrm{\Greektens{$\upzeta$}}$ represent the magneto-electric cross-susceptibilities. The noise polarization $\hat{\vec{P}}_{\textrm{N}}$ and noise magnetization $\hat{\vec{M}}_{\textrm{N}}$ form the noise current $\hat{\vec{j}}_{\textrm{N}}$
\begin{eqnarray}
\hat{\vec{j}}_{\textrm{N}} \left( \vec{r}, \omega \right) &=& - \mi \omega \hat{\vec{P}}_{\textrm{N}} \left( \vec{r}, \omega \right) + \overrightarrow{\vec{\nabla}} \vprod \hat{\vec{M}}_{\textrm{N}} \left( \vec{r}, \omega \right)\nonumber \\
&=& \begin{pmatrix} -\mi \omega, & \overrightarrow{\vec{\nabla}} \vprod \end{pmatrix} \sprod \begin{pmatrix} \hat{\vec{P}}_{\textrm{N}} \left( \vec{r}, \omega \right) \\ \hat{\vec{M}}_{\textrm{N}} \left( \vec{r}, \omega \right) \end{pmatrix}.
\label{eq:Noise Current Density1}
\end{eqnarray}
The noise polarization and noise magnetization can be expressed in terms of the creation and annihilation operators for the electric and the magnetic fields $\hat{\vec{f}}_{\textrm{e}}$, $\hat{\vec{f}}{}^{\dagger}_{\textrm{e}}$, $\hat{\vec{f}}_{\textrm{m}}$ and $\hat{\vec{f}}{}^{\dagger}_{\textrm{m}}$
\begin{equation}
\begin{pmatrix} \hat{\vec{P}}_{\textrm{N}} \\ \hat{\vec{M}}_{\textrm{N}} \end{pmatrix} = \sqrt{\frac{\hbar}{\pi}} \mathcal{R} \star \begin{pmatrix} \hat{\vec{f}}_{\textrm{e}} \\ \hat{\vec{f}}_{\textrm{m}} \end{pmatrix}.
\label{eq:Noise Polarization Magnetization}
\end{equation}
The Green's tensor $\tens{G}$ from Eq.~\eqref{eq:Electric Field Components} solves the respective Helmholtz equation
\begin{equation}
-\mu_0 \begin{pmatrix} - \mi \omega, & \overrightarrow{\vec{\nabla}} \vprod \end{pmatrix} \star \mathcal{M} \star \begin{pmatrix} \mi \omega \\ \overrightarrow{\vec{\nabla}} \vprod \end{pmatrix} \star \tens{G} = \textrm{\Greektens{$\updelta$}}
\end{equation}
with the matrix
\begin{equation}
\mathcal{M} = \begin{pmatrix} \epsilon_0 \left( \textrm{\Greektens{$\upepsilon$}} - \textrm{\Greektens{$\upxi$}} \star \textrm{\Greektens{$\upmu$}}^{-1} \star \textrm{\Greektens{$\upzeta$}} \right) & \frac{\textrm{\Greektens{$\upxi$}} \star \textrm{\Greektens{$\upmu$}}^{-1}}{Z_0} \\ \frac{\textrm{\Greektens{$\upmu$}}^{-1} \star \textrm{\Greektens{$\upzeta$}}}{Z_0} & -\frac{\textrm{\Greektens{$\upmu$}}^{-1}}{\mu_0} \end{pmatrix}.
\label{eq:M Matrix}
\end{equation}
The Helmholtz equation reduces to the standard form \cite{Buhmann_Book_1} if all cross-susceptibilities are set to $0$. The tensor $\mathcal{R}$ is related to the matrix $\mathcal{M}$ via
\begin{equation}
\mathcal{R} \star \mathcal{R}^{*T} = \Im \left[ \mathcal{M} \right].
\end{equation}
The conductivity matrix $\tens{Q}$ can also be expressed in terms of $\mathcal{M}$:
\begin{equation}
\tens{Q} = \frac{1}{\mi \omega} \begin{pmatrix} -\mi \omega, & \overrightarrow{\vec{\nabla}} \vprod \end{pmatrix} \star \left[ \mathcal{M} - \begin{pmatrix} \epsilon_0 & 0 \\ 0 & -\frac{1}{\mu_0} \end{pmatrix} \right] \star \begin{pmatrix} \mi \omega \\ - \vprod \overleftarrow{\vec{\nabla}} \end{pmatrix}.
\end{equation}
Calculations of the equations of motion of the creation and annihilation operators require the field Hamiltonian $\hat{H}_{\textrm{F}}$
\begin{equation}
\hat{H}_{\textrm{F}} = \sum\limits_{\lambda = \textrm{e,m}}{\int{\dif^3r \: \int\limits^{\infty}_0{\dif \omega \: \hbar \omega \hat{\vec{f}}{}^{\dagger}_{\lambda} \left( \vec{r}, \omega \right) \cdot \hat{\vec{f}}_{\lambda} \left( \vec{r}, \omega \right)}}}
\end{equation}
and the interaction Hamiltonian $\hat{H}_{\textrm{AF}}$ \eqref{eq:Interaction Hamiltonian}. Inserting Eqs.~\eqref{eq:Electric Field}, \eqref{eq:Electric Field Components}, \eqref{eq:Noise Current Density1} and \eqref{eq:Noise Polarization Magnetization} into Eq.~\eqref{eq:Interaction Hamiltonian} enables us to solve the linear and inhomogeneous differential equation of the field operators
\begin{multline}
\begin{pmatrix} \hat{\vec{f}}_{\textrm{e}} \left( \vec{r}, \omega, t \right) \\ \hat{\vec{f}}_{\textrm{m}} \left( \vec{r}, \omega, t \right) \end{pmatrix} =  \me^{-\mi \omega \left( t-t_0 \right)} \begin{pmatrix} \hat{\vec{f}}_{\textrm{e}} \left( \vec{r}, \omega \right) \\ \hat{\vec{f}}_{\textrm{m}} \left( \vec{r}, \omega \right) \end{pmatrix} + \frac{\mu_0 \omega}{\hbar} \sqrt{\frac{\hbar}{\pi}} \sum\limits_{m,n}{\int\limits^{t_0}_0{\dif t'}}\\
{{ \me^{-\mi \omega \left( t-t' \right)} \left[ \tens{G} \star \begin{pmatrix} -\mi \omega, & \overrightarrow{\vec{\nabla}} \: ' \times \end{pmatrix} \star \mathcal{R} \right]^{*\trans} \left( \vec{r}_A, \vec{r}, \omega \right) \sprod \vec{d}_{mn} \hat{A}_{mn}}},
\end{multline}
which can be inserted into Eqs.~\eqref{eq:Noise Polarization Magnetization}, \eqref{eq:Noise Current Density1} and \eqref{eq:Electric Field Components} again. After using Eq.~\eqref{eq:Identity Imaginary Part Green's Tensor} again, the final expression for $\hat{\vec{E}}$ yields
\begin{multline}
\hat{\vec{E}}\left( \vec{r}, \omega, t \right) = \me^{-\mi \omega \left( t-t_0 \right)} \hat{\vec{E}} \left( \vec{r}, \omega \right)\\
+ \mi \frac{\mu_0 \omega^2}{\pi} \sum\limits_{m,n}{\int\limits^{t}_{t_0}{\dif t' \: \me^{-\mi \omega \left( t-t' \right)} \Im \left[ \tens{G} \left( \vec{r}, \vec{r}_A, \omega \right) \right] \sprod \vec{d}_{mn} \hat{A}_{mn}}}
\label{eq:Results Components Electric Field2}
\end{multline}
and agrees perfectly with the result from Sec.~\ref{sec:General Derivation} \eqref{eq:Results Components Electric Field1}.

\section{Internal atomic dynamics: frequency shift and decay rate}
\label{sec:Internal Dynamics}
The internal atomic dynamics can be described by the Heisenberg equations of motion for the atomic flip operator
\begin{equation}
\dot{\hat{A}}_{mn} = \frac{1}{\mi \hbar} \left[ \hat A _{mn}, \hat H \right] = \frac{1}{\mi \hbar} \left[ \hat A_{mn}, \hat H_{\textrm{A}} \right] + \frac{1}{\mi \hbar} \left[ \hat A _{mn}, \hat H_{\textrm{AF}} \right]
\end{equation}
which includes only the atomic Hamiltonian $\hat{H}_{\textrm{A}}$ and the interaction Hamiltonian $\hat{H}_{\textrm{AF}}$ because the field Hamiltonian $\hat{H}_{\textrm{F}}$ commutes with the atomic flip operator. This approach follows the procedure for a reciprocal surface outlined in Ref.~\cite{Buhmann_Book_2} and is now extended to nonreciprocal media \cite{Buhmann_nonreciprocal:2012}.\\
This leads to
\begin{multline}
\dot{\hat{A}}_{mn} = \mi \omega_{mn} \hat A _{mn} + \frac{\mi}{\hbar} \sum\limits_k{\int\limits^{\infty}_0{\dif \omega}}\\
\left[ \left( \hat A _{mk} \vec{d}_{nk} - \hat A_{kn} \vec{d}_{km} \right) \cdot \hat{\vec{E}} \left( \vec{r}_A, \omega \right) \right.\\
+\left. \hat{\vec{E}}{}^{\dagger} \left( \vec{r}_A, \omega \right) \cdot \left( \vec{d}_{nk} \hat A_{mk} - \vec{d}_{km} \hat A _{kn} \right) \right].
\label{eq:Differential Equation Atomic Flip Operator}
\end{multline}
$\hat A _{mn}$ is dominated by oscillations with frequencies $\tilde{\omega}_{mn} = \omega_{mn} + \delta \omega_{mn}$, where $\omega_{mn}$ is the atom's eigenfrequency and $\delta \omega_{mn}$ is the shift owing to interaction with nearby material bodies (Casimir--Polder shift). The electric field is given in Eqs.~\eqref{eq:Results Components Electric Field1} or \eqref{eq:Results Components Electric Field2}. The time-integral in the electric field can be formally evaluated in the Markov approximation where we neglect the slow non-oscillatory dynamics of the atomic flip operator $\hat{A}_{mn}$ during the time interval $t_0 \leq t' \leq t$ and set $\hat A_{mn} \left( t' \right) \simeq \exp{ \left[ \mi \tilde{\omega}_{mn} \left( t'-t \right) \right]} \hat A_{mn} \left( t \right)$, where we have anticipated the result $\tilde{\omega}_{mn} = -\tilde{\omega}_{nm}$. In the long time limit $t \rightarrow \infty$ the time integral reduces to $\hat A_{mn} \left( t \right) \int^{t}_{t_0}{\dif t' \: \exp{ \left[ -\mi \left( \omega-\tilde{\omega}_{nm} \right) \left( t-t' \right) \right]}} \simeq A_{mn} \left( t \right) \left[ \pi \delta \left( \omega - \tilde{\omega}_{nm} \right) - \mi \mathcal{P}/ \left( \omega - \tilde{\omega}_{nm} \right)\right]$, where $\mathcal{P}$ is the Cauchy principle value and the limits of the frequency integral will lead to the appearance of the Heaviside step-function $\Theta$.\\
By defining the coefficient
\begin{multline}
\vec{C}_{mn} = \frac{\mu_0}{\hbar} \Theta \left( \tilde{\omega}_{nm} \right) \tilde{\omega}^2_{nm} \Im \left[ \tens{G} \left( \vec{r}_{\textrm{A}}, \vec{r}_{\textrm{A}}, \tilde{\omega}_{nm} \right) \right] \sprod \vec{d}_{mn}\\ - \mi \frac{\mu_0}{\pi \hbar} \mathcal{P} \int\limits^{\infty}_0{ \dif \omega \frac{1}{\omega - \tilde{\omega}_{nm}} \omega^2 \Im \left[ \tens{G} \left( \vec{r}_{\textrm{A}}, \vec{r}_{\textrm{A}}, \omega \right) \right] \sprod \vec{d}_{mn}},
\label{eq:Coefficient C}
\end{multline}
Eq.~\eqref{eq:Differential Equation Atomic Flip Operator} can be cast into the form
\begin{multline}
\dot{\hat{A}}_{mn} \left( t \right) = \mi \omega_{mn} \hat{A}_{nm} \left( t \right)\\
+ \frac{\mi}{\hbar} \sum_k \int\limits^{\infty}_0 \dif \omega \left\{ \me^{- \mi \omega \left( t-t_0 \right)} \left[ \hat{A}_{mk} \left( t \right) \vec{d}_{nk} - \hat{A}_{kn} \left( t\right) \vec{d}_{km} \right] \sprod \hat{\vec{E}} \left( \vec{r}_{\textrm{A}}, \omega \right) \right.\\
\left. + \me^{\mi \omega \left( t-t_0 \right)} \hat{\vec{E}}{}^{\dagger} \left( \vec{r}_{\textrm{A}}, \omega \right) \sprod \left[ \vec{d}_{nk} \hat{A}_{mk} \left( t \right) - \vec{d}_{km} \hat{A}_{kn} \left( t\right) \right] \right\}\\
-\sum\limits_{k,l} \left[ \vec{d}_{nk} \sprod \vec{C}_{kl} \hat{A}_{ml} \left( t \right) - \vec{d}_{km} \sprod \vec{C}_{nl} \hat{A}_{kl} \left( t\right) \right]\\
+ \sum \limits_{k,l} \left[ \vec{d}_{nk} \sprod \vec{C}^*_{ml} \hat{A}_{lk} \left( t \right) - \vec{d}_{km} \sprod \vec{C}^*_{kl} \hat{A}_{ln} \left( t \right) \right],\\
\label{eq:Differential Equation Atomic Flip Operator2}
\end{multline}
where we have used the identity $\Im \left[ \tens{G}^* \left( \vec{r}_{\textrm{A}}, \vec{r}_{\textrm{A}}, \omega \right) \right] = \Im \left[ \tens{G}^{\trans} \left( \vec{r}_{\textrm{A}}, \vec{r}_{\textrm{A}}, \omega \right) \right]$, which can be derived from Eq.~\eqref{eq:Imaginary Part}.\\
Next, we take expectation values of Eq.~\eqref{eq:Differential Equation Atomic Flip Operator2} and assume the electromagnetic field to be prepared in its ground state at initial time $t_0$ which implies $\hat{\vec{E}} \left( \vec{r}, \omega \right) |\left\{ 0 \right\} \rangle = \veczero$. Therefore, the free terms of the electric field $\hat{\vec{E}} \left( \vec{r}, \omega \right)$ and $\hat{\vec{E}}{}^{\dagger} \left( \vec{r}, \omega \right)$ do not contribute to the dynamics of the average atomic flip operator's value and are discarded.\\
Since we assume the atom to be free of quasi-degenerate transitions, the set of differential equations for the atomic flip operator's expectation value can be decoupled. Moreover, the atom is unpolarized in each of its energy eigenstates, $\hat{\vec{d}}_{nn} = 0$, which is guaranteed by atomic selection rules \cite{Buhmann_Book_2}. As a result of these assumptions, the fast-oscillating off-diagonal flip operators decouple from the non-oscillating diagonal ones as well as from each other \cite{Buhmann_Book_2}.\\
By making use of Eq.~\eqref{eq:Imaginary Part} we find that the two terms $\vec{d}_{nk} \cdot \Im \left[ \tens{G} \left( \vec{r}_A, \vec{r}_A, \omega \right) \right] \cdot \vec{d}_{kn}= \mathrm{Im} \left[ \vec{d}_{nk} \cdot \tens{G} \left( \vec{r}_A, \vec{r}_A, \omega \right) \cdot \vec{d}_{kn} \right]$ and $\vec{d}_{kn} \cdot \Im \left[ \tens{G}^{\trans} \left( \vec{r}_A, \vec{r}_A, \omega \right) \right] \cdot \vec{d}_{nk}=\mathrm{Im} \left[ \vec{d}_{nk} \cdot \tens{G} \left( \vec{r}_A, \vec{r}_A, \omega \right) \cdot \vec{d}_{kn} \right]$ are equal and real.\\
With the help of these relations we identify the decay rate
\begin{equation}
\Gamma_{nk} = \frac{2 \mu_0}{\hbar} \tilde{\omega}^2_{nk} \mathrm{Im} \left[ \vec{d}_{nk} \cdot \tens{G} \left( \vec{r}_A, \vec{r}_A, \tilde{\omega}_{nk} \right) \cdot \vec{d}_{kn} \right]
\label{eq:Decay Rate}
\end{equation}
and the frequency shift
\begin{multline}
\delta \omega_{nk} =\\ -\frac{\mu_0}{\pi \hbar} \mathcal{P} \int\limits^{\infty}_0{\dif \omega \: \frac{1}{\omega - \tilde{\omega}_{nk}}\omega^2 \mathrm{Im} \left[ \vec{d}_{nk} \cdot \tens{G}^{(1)} \left( \vec{r}_A, \vec{r}_A, \omega \right) \cdot \vec{d}_{kn} \right]}.
\label{eq:Frequency Shift}
\end{multline}
Here, the Green's tensor $\tens{G}$ has been split into a bulk part $\tens{G}^{(0)}$ and a scattering part $\tens{G}^{(1)}$. The Lamb shift due to the free-space Green's tensor $\tens{G}^{(0)}$ is already included in the transition frequency $\omega_{mn}$, which refers solely to the atom and does not take the material properties of surrounding matter into account. The remaining frequency shift stems from the presence of electromagnetic bodies around the atom.\\
Finally, the expectation value for the atomic flip operator for the non-diagonal terms yields 
\begin{eqnarray}
\langle \dot{\hat{A}}_{mn} \left( t \right) \rangle &=&\mi \omega_{mn} \langle \hat A_{mn} \left( t \right) \rangle \nonumber\\
& &+ \sum\limits_k{\left( -\frac{1}{2} \Gamma_{nk} - \mi \delta \omega_{nk} \right)} \langle \hat A_{mn} \left( t \right) \rangle \nonumber\\
& &+ \sum\limits_k{\left( -\frac{1}{2} \Gamma_{mk} + \mi \delta \omega_{mk} \right)} \langle \hat A_{mn} \left( t \right) \rangle.
\label{eq:Expectation Value Atomic Flip Operator}
\end{eqnarray}
We define $\delta \omega_n = \sum\limits_k{\delta \omega_{nk}}$ and $\Gamma_{n} = \sum\limits_{k<n}{\Gamma_{nk}}$ --- the $\Theta$-function in Eq.~\eqref{eq:Coefficient C} determines the order of summation indices --- and the shifted transition frequency as
\begin{equation}
\tilde{\omega}_{mn} = \omega_{mn} + \delta \omega_m - \delta \omega_n,
\end{equation}
which verifies our previous assumption $\tilde{\omega}_{mn} = -\tilde{\omega}_{nm}$. Thus Eq.~\eqref{eq:Expectation Value Atomic Flip Operator} for the diagonal terms has the simple form
\begin{equation}
\langle \dot{\hat{A}}_{nn} \left( t \right) \rangle = - \Gamma_n \langle \hat{A}_{nn} \left( t \right) \rangle + \sum\limits_{k>n}{\Gamma_{kn} \langle \hat{A}_{kk} \left( t \right) \rangle}.
\end{equation}
Since the shifted frequency $\tilde{\omega}_{nk}$ appears in $\delta \omega_{nk}$ itself, the frequency shift is given as a self-consistent result from the implicit equation.\\ 
The complex frequency shift \eqref{eq:Frequency Shift} can be simplified further by making use of the definition of the imaginary part \eqref{eq:Imaginary Part}, the Schwarz principle which is still valid for nonreciprocal media
\begin{equation}
\tens{G}^* \left( \vec{r}_A, \vec{r}_A, \omega \right) = \tens{G} \left( \vec{r}_A, \vec{r}_A, -\omega^* \right)
\end{equation}
and a substitution $\omega \rightarrow - \omega$ in the second integral having its origin in Eq.~\eqref{eq:Imaginary Part}. The integral contours along the positive and negative real axes have one pole each and are evaluated in the complex plane. The path along the quarter circle does not give a contribution because $\lim_{|\omega| \rightarrow 0} \tens{G}^{(1)} \left( \vec{r}, \vec{r} \: ', \omega \right) \omega^2/c^2 = 0$. The part along the imaginary axis leads to the nonresonant frequency shift
\begin{multline}
\delta \omega^{\textrm{nres}}_{nk} = \frac{\mu_0}{\pi \hbar} \int\limits^{\infty}_0{\dif \xi \: \frac{\xi^3}{\xi^2 + \tilde{\omega}^2_{nk}} \mathrm{Im} \left[ \vec{d}_{nk} \cdot \tens{G}^{(1)}  \left( \vec{r}_A, \vec{r}_A, \mi \xi \right) \cdot \vec{d}_{kn} \right]}\\
- \frac{\mu_0}{\pi \hbar} \int\limits^{\infty}_0{\dif \xi \: \frac{\xi^2 \tilde{\omega}_{nk}}{\xi^2 + \tilde{\omega}^2_{nk}} \mathrm{Re} \left[ \vec{d}_{nk} \cdot \tens{G}^{(1)}  \left( \vec{r}_A, \vec{r}_A, \mi \xi \right) \cdot \vec{d}_{kn}\right]}
\label{eq:Frequency Shift Nonresonant Part}
\end{multline}
with a Green's function $\tens{G}$ with imaginary frequency $\omega \rightarrow i \xi$. This expression resembles the frequency shift of the Casimir--Polder force for an atom in its ground state \cite{Buhmann_Book_1}. It comes from the exchange of virtual photons between the atom and the material body. This entirely quantum mechanical interpretation can be extended for an atom in an arbitrary state. The matrix-vector product of the Green's tensor and the dipole moments is real for a reciprocal medium and therefore only the second contribution remains in this case.\\
The evaluation of the poles gives the resonant contribution associated with real-photon emission and a real frequency expression $\tilde{\omega}_{nk}$. 
\begin{equation}
\delta \omega^{\textrm{res}}_{nk} = -\frac{\mu_0}{\hbar} \tilde{\omega}^2_{nk} \mathrm{Re} \left[ \vec{d}_{nk} \cdot \tens{G}^{(1)}  \left( \vec{r}_A, \vec{r}_A, \tilde{\omega}_{nk} \right) \cdot \vec{d}_{kn} \right].
\label{eq:Frequency Shift Resonant Part}
\end{equation}
In case of the resonant frequency shift, the Green's tensor $\tens{G}$ in Eq.~\eqref{eq:Frequency Shift Resonant Part} contains discrete frequencies for the real atomic transitions to a lower energy state, which can only occur for excited atoms and is related to real exchange photons.\\
The sum of the resonant/nonresonant frequency shifts $\delta \omega_{nk}$ over all indices $k$ can be identified with the position-dependent resonant/nonresonant Casimir--Polder potential. Its derivative with respect to position is the Casimir--Polder force between the atom and the nonreciprocal medium, which is caused by the atom's level-shift due to the body's presence.

\section{Applications and results}
\label{sec:Applications and Results}
\begin{figure}[!ht]
\centerline{\includegraphics[width=\columnwidth]{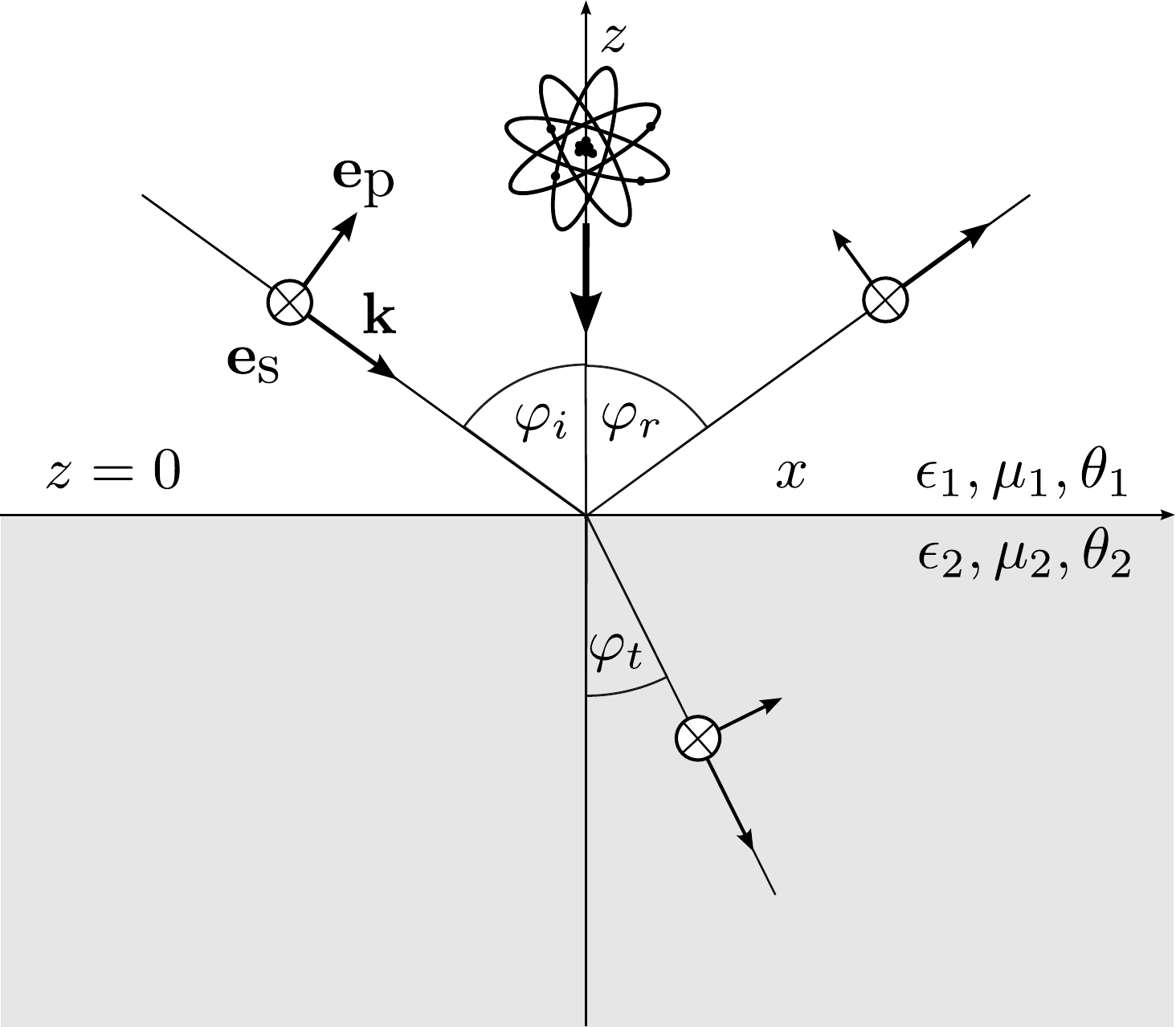}}
\caption{Sketch of an atom in front of a medium with electric, magnetic properties and an axion coupling. The directions of incoming parallely polarized light $\vec{e}_{\textrm{p}}$ and perpendicularly polarized light $\vec{e}_{\textrm{s}}$ are shown.}
\label{fig:Casimir_Polder_Paper}
\end{figure}

Having derived expressions for the atomic rate of spontaneous decay \eqref{eq:Decay Rate} and nonresonant/resonant frequency shifts \eqref{eq:Frequency Shift Nonresonant Part}, \eqref{eq:Frequency Shift Resonant Part}, we contrast a perfectly reflecting nonreciprocal mirror to a perfectly conducting mirror. Afterward we compare this to a topological insulator from Ref.~\cite{Crosse:2015}. Fig.~\ref{fig:Casimir_Polder_Paper} shows a sketch of an atom in front of a medium having electric, magnetic properties and an axion coupling.
The scattering part of the Green's tensor $\tens{G}^{(1)}$ of a single planar surface has the form \cite{Crosse:2015}
\begin{multline}
\tens{G}^{(1)} \left( \vec{r}, \vec{r} \: ', \omega \right) = \frac{\mi}{8 \pi^2}\int\dif^2 k^{\parallel} {\frac{1}{k^{\perp}} \: \sum\limits_{\sigma=\textrm{s},\textrm{p}}{\sum\limits_{\sigma'=\textrm{s},\textrm{p}}}}\\
r_{\sigma,\sigma'} \vec{e}_{\sigma+} \tprod \vec{e}_{\sigma'-} \me^{\mi \vec{k}^{\parallel} \left( \vec{r}-\vec{r} \: ' \right)} \me^{\mi k^{\perp} \left( z + z \: ' \right)},
\label{eq:Green's Tensor Scattering Part} 
\end{multline}
with the two unit vectors $\vec{e}_{\sigma+}$ and $\vec{e}_{\sigma'-}$, representing the polarizations of incident $(\sigma')$ and reflected waves $(\sigma)$. The reflective coefficient $r_{\sigma,\sigma'}$ takes the mixing of the incoming and outgoing polarizations $\sigma'$ and $\sigma$ into account. The indices $\textrm{p}$ and $\textrm{s}$ refer to parallel or perpendicular polarization. $\vec{k}^{\parallel}$ representing the parallel component of the wave vector, $\vec{k}^{\perp}$ its perpendicular component and $z$ is the vertical distance to the surface.\\
According to Curie's principle a system consisting of a crystal and an external influence, each having a specific symmetry, only maintains the symmetries that are shared by both the crystal and the external influence \cite{Curie:1894}. Hence our choice of dipole moments must be such that the atom is sensitive to the violated time-reversal symmetry of a perfectly reflecting nonreciprocal mirror. To study possible effects of nonreciprocity, we assume circularly polarized dipole moments
\begin{equation}
\vec{d}_{10} = \frac{d}{\sqrt{2}} \begin{pmatrix} 1 \\ \mi \\ 0 \end{pmatrix}, \: \vec{d}_{01} = \frac{d}{\sqrt{2}} \begin{pmatrix} 1 \\ -\mi \\ 0 \end{pmatrix}
\label{eq:Circularly Polarized Dipole}
\end{equation}
which are not invariant if the direction of time is reversed $t \rightarrow -t$.

\subsection{Perfectly Conducting Mirror}
\label{sec:Perfectly Reflecting Mirror}
\begin{figure}[!ht]
\centerline{\includegraphics[width=\columnwidth]{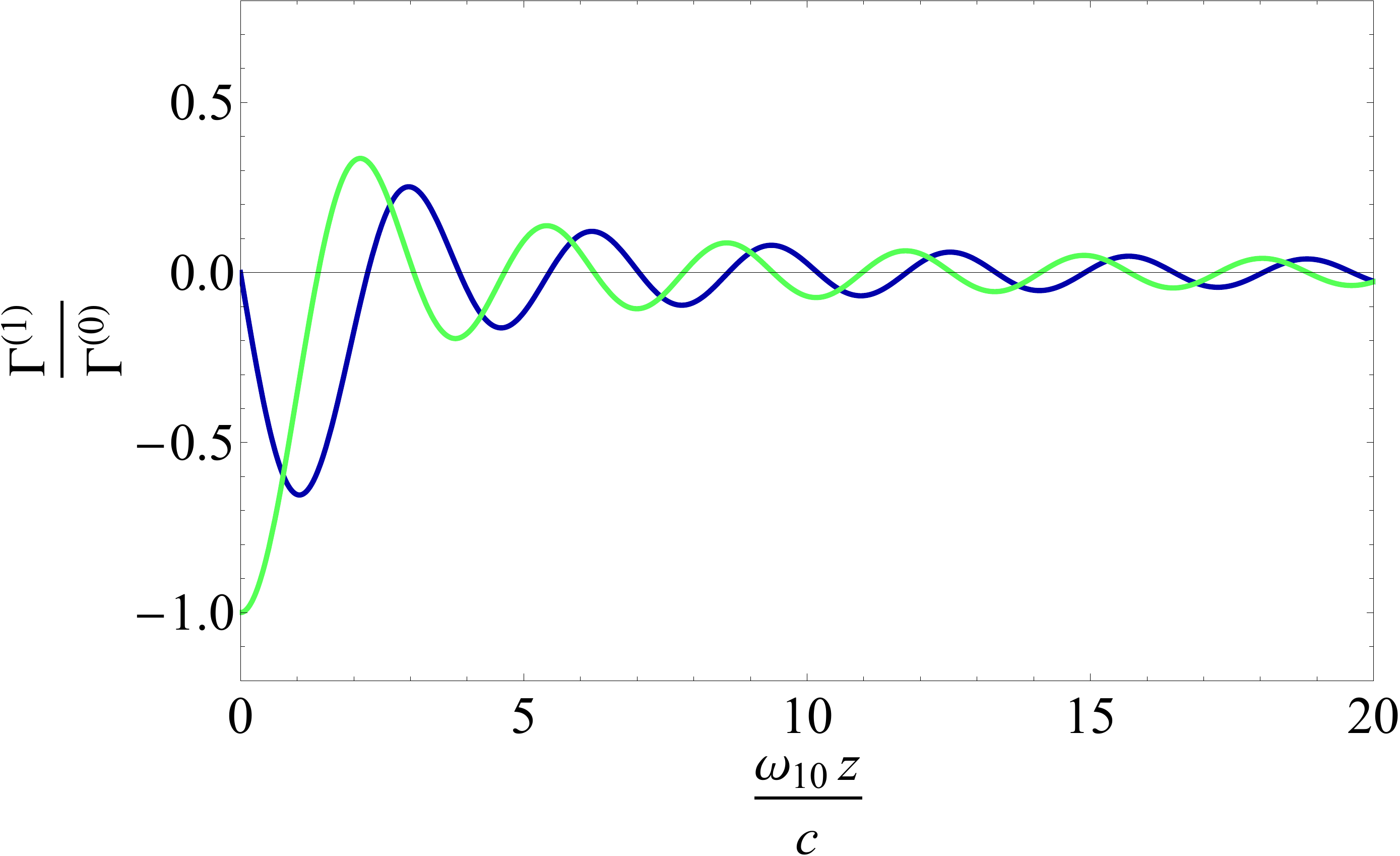}}
\caption{Atomic decay rates $\Gamma^{(1)}$ scaled by the free-space decay rate $\Gamma^{(0)}$ \eqref{eq:Free Space Decay Rate} for a circularly polarized two-level atomic dipole in front of a perfectly conducting mirror ($\textcolor{LimeGreen}{\hdashrule[0.5ex][x]{0.6cm}{1pt}{}}$) and a perfectly reflecting nonreciprocal mirror ($\textcolor{Blue}{\hdashrule[0.5ex][x]{0.6cm}{1pt}{}}$) or .}
\label{fig:PlotGamma}
\end{figure}

\begin{figure}[!ht]
\centerline{\includegraphics[width=\columnwidth]{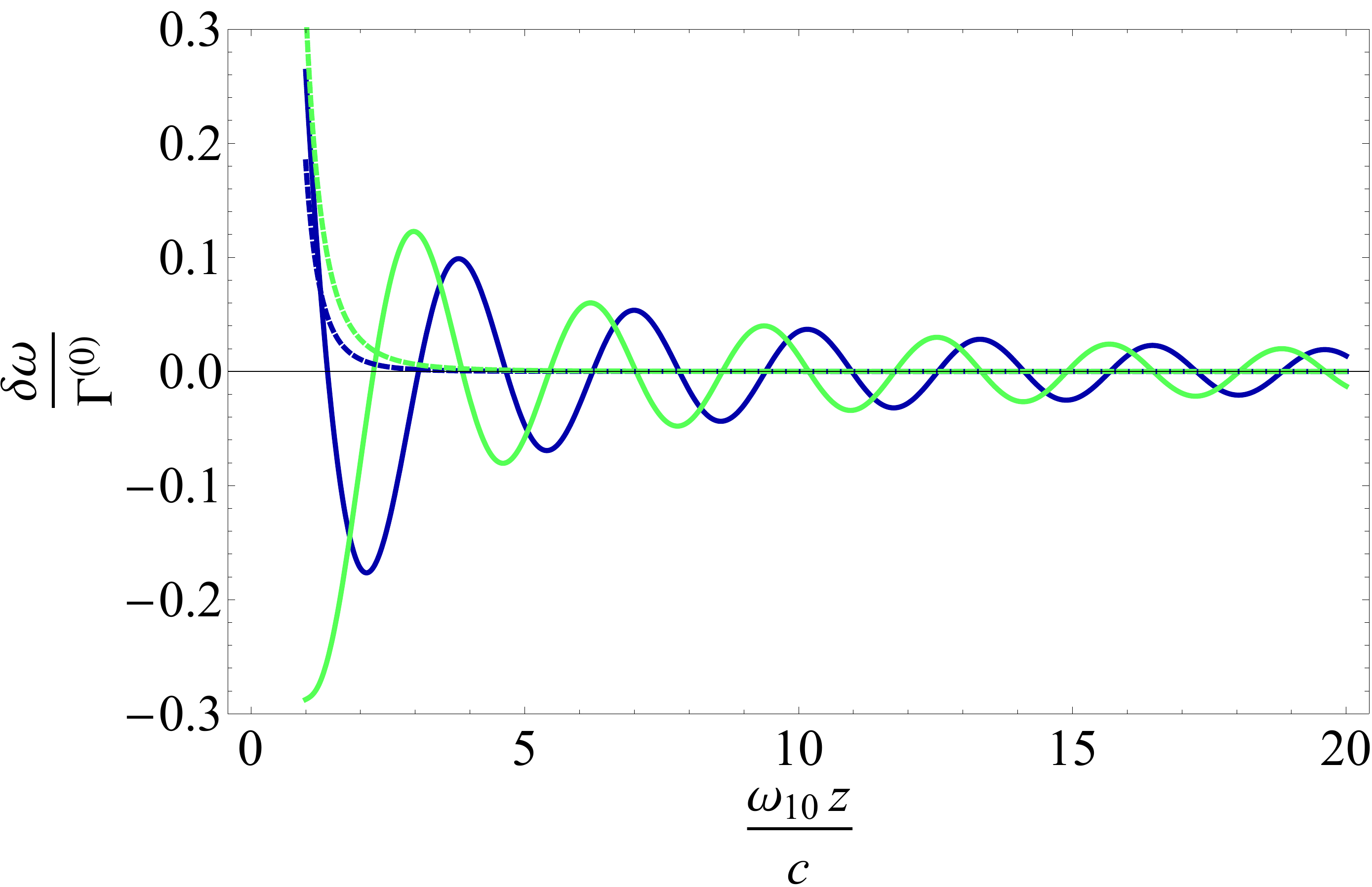}}
\caption{Frequency shifts $\delta \omega$ scaled by the free-space decay rate $\Gamma^{(0)}$ \eqref{eq:Free Space Decay Rate} for a circularly polarized two-level atomic dipole in front of a perfectly reflecting nonreciprocal mirror and a perfectly conducting mirror. The resonant frequency shift $\delta \omega^{\textrm{res}}$ of the perfectly conducting mirror ($\textcolor{LimeGreen}{\hdashrule[0.5ex][x]{0.6cm}{1pt}{}}$) and the resonant frequency shift of the perfectly reflecting nonreciprocal mirror ($\textcolor{Blue}{\hdashrule[0.5ex][x]{0.6cm}{1pt}{}}$) show oscillations. The nonresonant frequency shift $\delta \omega^{\textrm{nres}}$ of the perfectly conducting mirror ($\textcolor{LimeGreen}{\hdashrule[0.5ex][x]{0.6cm}{1pt}{1mm 1pt}}$) and the perfectly reflecting nonreciprocal mirror ($\textcolor{Blue}{\hdashrule[0.5ex][x]{0.6cm}{1pt}{1mm 1pt}}$) decay monotonously with distance.} 
\label{fig:PlotOmega}
\end{figure}

\begin{figure}[!ht]
\centerline{\includegraphics[width=\columnwidth]{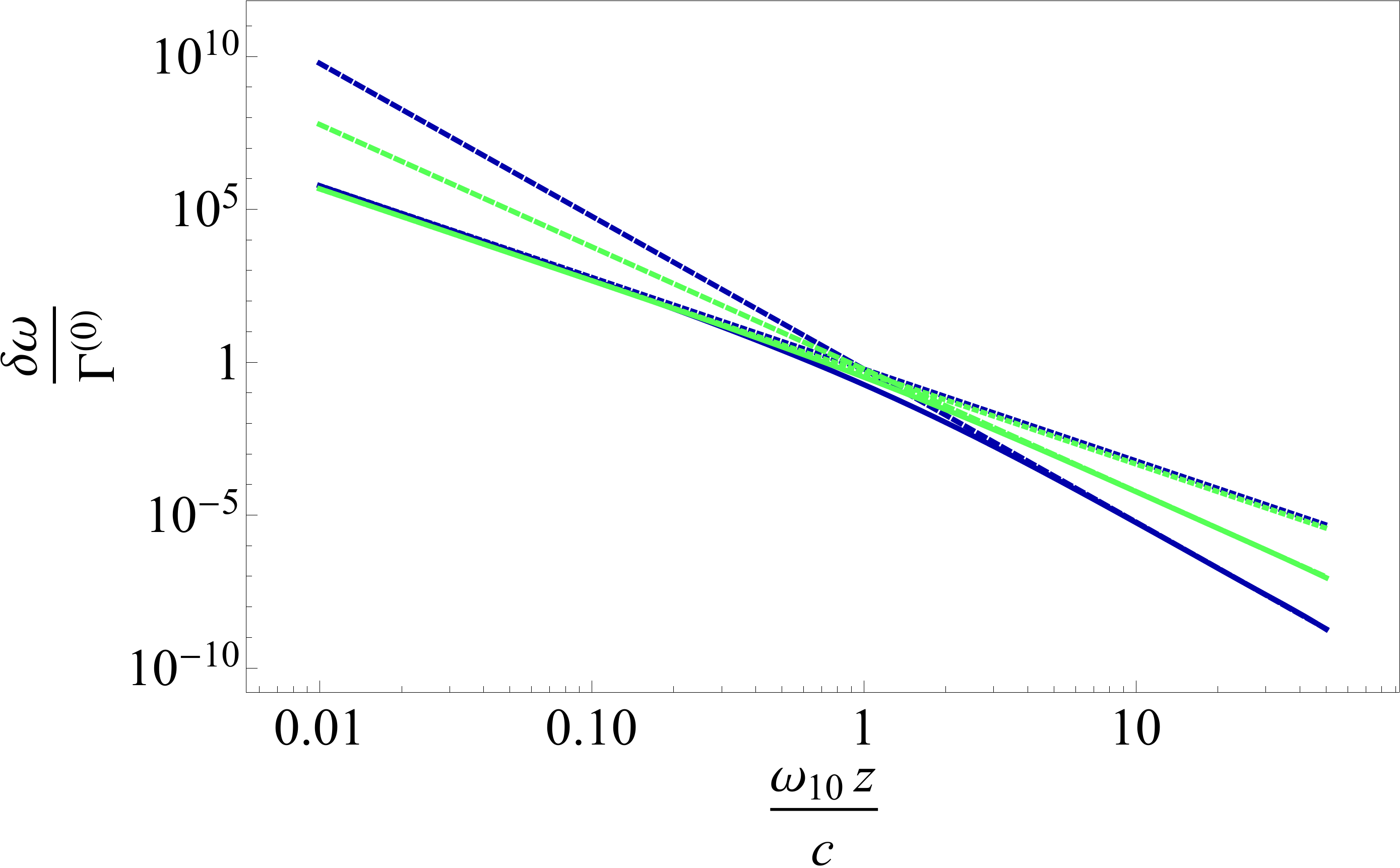}}
\caption{Double logarithmic plot for the nonresonant frequency shift $\delta \omega^{\textrm{nres}}$ of the perfectly conducting mirror ($\textcolor{LimeGreen}{\hdashrule[0.5ex][x]{0.6cm}{1pt}{}}$), its retarded limit ($\textcolor{LimeGreen}{\hdashrule[0.5ex][x]{0.6cm}{1pt}{1mm 1pt}}$) and its nonretarded limit ($\textcolor{LimeGreen}{\hdashrule[0.5ex][x]{0.6cm}{1pt}{0.2mm}}$). The perfectly reflecting nonreciprocal mirror ($\textcolor{Blue}{\hdashrule[0.5ex][x]{0.6cm}{1pt}{}}$), its retarded limit ($\textcolor{Blue}{\hdashrule[0.5ex][x]{0.6cm}{1pt}{1mm 1pt}}$) and its nonretarded limit ($\textcolor{Blue}{\hdashrule[0.5ex][x]{0.6cm}{1pt}{0.2mm}}$) are depicted in the same figure.}
\label{fig:PlotLogLog}
\end{figure}

Let us first investigate the atomic decay rate \eqref{eq:Decay Rate} and the nonresonant/resonant frequency shift \eqref{eq:Frequency Shift Nonresonant Part}, \eqref{eq:Frequency Shift Resonant Part} for a perfectly conducting mirror. The energy shift of a hydrogen atom between two conducting plates has been studied in Ref.~\cite{Barton:1970, Barton:1979} and one can obtain the interaction between an atom and a single plate if one plate is shifted to infinity. Ref.~\cite{Hinds_Book} shows the radiative decay rate of an atom in front of a perfect mirror, where the dipole is either parallel or perpendicular to the mirror. These approaches are based on perturbation theory.\\
The reflective coefficients for a perfectly conducting mirror are $r_{\textrm{p},\textrm{p}}=1$, $r_{\textrm{s},\textrm{s}}=-1$ and $r_{\textrm{s},\textrm{p}}=r_{\textrm{p},\textrm{s}}=0$. In this case the Green's tensor \eqref{eq:Green's Tensor Scattering Part} contains only diagonal terms $\tens{G}^{(1)}_{xx} = \tens{G}^{(1)}_{yy}$ with
\begin{equation}
\tens{G}^{(1)}_{xx} \left( \vec{r}, \vec{r}, \omega \right) = \left(-\frac{1}{8 \pi z} -\mi \frac{c}{16 \pi \omega z^2} + \frac{c^2}{32 \pi \omega^2 z^3} \right) \me^{\frac{2 \mi \omega z}{c}}.
\end{equation}
The nondiagonal elements of the Green's tensor vanish. The atomic decay rate \eqref{eq:Decay Rate} for circularly polarized dipole moments \eqref{eq:Circularly Polarized Dipole} hence reads
\begin{multline}
\Gamma^{(1)}_{10} = \frac{\mu_0 \tilde{\omega}^2_{10} d^2}{4 \pi \hbar} \left[ -\frac{1}{z} \sin \left( \frac{2 \tilde{\omega}_{10} z}{c} \right) \right.\\
\left. - \frac{c}{2 \tilde{\omega}_{10} z^2} \cos \left( \frac{2 \tilde{\omega}_{10} z}{c} \right) + \frac{c^2}{4 \tilde{\omega}^2_{10} z^3} \sin \left( \frac{\tilde{2 \omega}_{10} z}{c} \right) \right].
\label{eq:Decay Rate Perfectly Reflecting Mirror}
\end{multline}
Fig.~\ref{fig:PlotGamma} shows the atomic decay rate \eqref{eq:Decay Rate Perfectly Reflecting Mirror} scaled by the free-space decay rate
\begin{equation}
\Gamma^{(0)}_{10}=\frac{\mu_0 \tilde{\omega}^3_{10} d^2}{3 \pi \hbar c}.
\label{eq:Free Space Decay Rate}
\end{equation}
Moreover we study the asymptotic behavior of the decay rate and distinguish between the retarded limit $( \tilde{\omega}_{10} z/c \gg 1 )$ and the nonretarded limit $( \tilde{\omega}_{10} z/c \ll 1 )$. The decay rate decays asymptotically in the retarded limit with $-\left[ \mu_0 \tilde{\omega}^2_{10} d^2 \sin \left( 2 \tilde{\omega}_{10} z/c \right) \right] / \left[ 4 \pi \hbar z \right]$.\\
At $z=0$, in the nonretarded limit, the decay rate has a value of $-\Gamma^{(0)}_{10}$ (Fig.~\ref{fig:PlotGamma}). The total decay rate $\Gamma_{10}$ of a dipole parallel to a perfectly conducting mirror is a sum of the free-space part $\Gamma^{(0)}_{10}$ and the body-induced part $\Gamma^{(1)}_{10}$ and is equal to $\Gamma_{10} = \Gamma^{(0)}_{10} + \Gamma^{(1)}_{10}=0$ on the surface of the mirror at $z=0$. This can be explained by an image dipole with equal strength and opposite direction induced by the original one so that  the two dipoles cancel, leading to vanishing radiative decay.\\
The frequency shift is composed of a resonant and a nonresonant contribution
\begin{multline}
\delta \omega_{10} = \delta \omega^{\textrm{res}}_{10} + \delta \omega^{\textrm{nres}}_{10}\\
=\frac{\mu_0 \tilde{\omega}^2_{10} d^2}{8 \pi \hbar}  \left[ \frac{1}{z} \cos \left( \frac{2 \tilde{\omega}_{10} z}{c} \right) - \frac{c}{2 \tilde{\omega}_{10} z^2} \sin \left( \frac{2 \tilde{\omega}_{10} z}{c} \right) \right.\\
\left. - \frac{c^2}{4 \tilde{\omega}^2_{10} z^3} \cos \left( \frac{2 \tilde{\omega}_{10} z}{c} \right) \right]\\
+ \frac{\mu_0 d^2}{8 \pi^2 \hbar} \int\limits^{\infty}_0{\dif \xi \: \frac{\tilde{\omega}_{10} \xi^2}{\tilde{\omega}^2_{10} + \xi^2} \left( \frac{1}{z} + \frac{c}{2 \xi z^2} + \frac{c^2}{4 \xi^2 z^3} \right) \me^{- \frac{2 \xi z}{c}}},
\label{eq:Frequency Shift Perfectly Reflecting Mirror}
\end{multline}
which are shown in Fig.~\ref{fig:PlotOmega}.
The retarded and nonretarded limits of the nonresonant frequency shift \eqref{eq:Frequency Shift Perfectly Reflecting Mirror} read
\begin{align}
\begin{array}{lll}
&\delta \omega^{\textrm{nres}}_{10} &= \begin{cases} \displaystyle{\frac{d^2 c}{16 \pi^2 \epsilon_0 \hbar \tilde{\omega}_{10} z^4}}, & \displaystyle{\frac{\tilde{\omega}_{10} z}{c} \gg 1}, \\[2mm] \displaystyle{\frac{d^2}{64 \pi \epsilon_0 \hbar z^3}}, & \displaystyle{\frac{\tilde{\omega}_{10} z}{c} \ll 1} \end{cases}
\end{array}
\end{align}
and are depicted in a double logarithmic plot in Fig.~\ref{fig:PlotLogLog}.
The asymptotic limits of the resonant frequency shift \eqref{eq:Frequency Shift Perfectly Reflecting Mirror} read
\begin{align}
\begin{array}{lll}
&\delta \omega^{\textrm{res}}_{10} &= \begin{cases} \displaystyle{\frac{\mu_0 \tilde{\omega}^2_{10} d^2}{8 \pi \hbar z} \cos \left( \frac{2 \tilde{\omega}_{10} z}{c} \right)}, & \displaystyle{\frac{\tilde{\omega}_{10} z}{c} \gg 1}, \\[2mm] \displaystyle{-\frac{d^2}{32 \pi \epsilon_0 \hbar z^3}}, & \displaystyle{\frac{\tilde{\omega}_{10} z}{c} \ll 1} \end{cases}\\[2mm]
\end{array}
\end{align}
and are shown in Fig.~\ref{fig:PlotOmega} as well.

\subsection{Perfectly Reflecting Nonreciprocal Mirror}
\label{sec:Perfect Nonreciprocal Medium}
The reflection coefficients for incoming perpendicular/parallel polarization and outgoing perpendicular/parallel polarization $r_{\textrm{s},\textrm{s}}$ and $r_{\textrm{p},\textrm{p}}$ are set equal to $0$, whereas the mixing terms $r_{\textrm{s,p}}$ and $r_{\textrm{p,s}}$ can be chosen to be either $1$ or $-1$ thus generating a perfectly reflecting nonreciprocal mirror. In this section, we restrict ourselves to the case $r_{\textrm{s,p}} = r_{\textrm{p,s}} = -1$. The Green's tensor \eqref{eq:Green's Tensor Scattering Part} is anti-symmetric under these conditions: $\tens{G}^{(1) \textrm{T}} \left( \vec{r}, \vec{r} \: ', \omega \right) = -\tens{G}^{(1)} \left( \vec{r} \: ' , \vec{r}, \omega \right)$. Thus the diagonal terms of the Green's tensor vanish and only the non-diagonal terms remain. By interchanging the indices of the non-diagonal terms, $\tens{G}^{(1)}_{xz} \left( \vec{r}, \vec{r} \: ' \right) = \tens{G}^{(1)}_{zx} \left( \vec{r}, \vec{r} \: ' \right)$ and $\tens{G}^{(1)}_{yz} \left( \vec{r}, \vec{r} \: ' \right) = \tens{G}^{(1)}_{zy} \left( \vec{r}, \vec{r} \: ' \right)$ keep their signs, whereas $\tens{G}^{(1)}_{xy} \left( \vec{r}, \vec{r} \: ' \right) = -\tens{G}^{(1)}_{yx} \left( \vec{r}, \vec{r} \: ' \right)$ shows a sign change. This behavior is exactly opposite if the arguments of the non-diagonal terms are interchanged. Therefore $\tens{G}^{(1)}_{xz}$ and $\tens{G}^{(1)}_{yz}$ have to vanish by setting $\vec{r} = \vec{r} \: '$ and only $\tens{G}^{(1)}_{xy} = -\tens{G}^{(1)}_{yx}$ has finite values. The final result after integrating yields
\begin{equation}
\tens{G}^{(1)}_{xy} \left( \vec{r}, \vec{r}, \omega \right)= \left( -\frac{1}{8 \pi z} - \mi \frac{c}{16 \pi \omega z^2} \right) \me^{\frac{2 \mi \omega z}{c}}.
\end{equation}
By using the circularly polarized dipole moments \eqref{eq:Circularly Polarized Dipole} we obtain for the atomic decay rate
\begin{multline}
\Gamma^{(1)}_{10} = \frac{\mu_0 \tilde{\omega}^2_{10} d^2}{4 \pi \hbar} \left[ \frac{1}{z} \cos \left( \frac{2 \tilde{\omega}_{10} z}{c} \right) \right.\\
\left. - \frac{c}{2 \tilde{\omega}_{10} z^2} \sin \left( \frac{2 \tilde{\omega}_{10} z}{c} \right) \right],
\label{eq:Decay Rate Perfect Nonreciprocal Medium}
\end{multline}
which is shown in Fig.~\ref{fig:PlotGamma}. The decay rate of a perfectly reflecting nonreciprocal mirror is equal to $0$ for small values of $\tilde{\omega}_{10} z/c$ (nonretarded limit). The function decays asymptotically in the retarded limit with $[\mu_0 \tilde{\omega}^2_{10} d^2 \cos \left( 2 \tilde{\omega}_{10} z/c \right)] / [4 \pi \hbar z]$.\\
The frequency shift is shown in Fig.~\ref{fig:PlotOmega} and consists of a resonant and a nonresonant contribution
\begin{multline}
\delta \omega_{10} = \delta \omega^{\textrm{res}}_{10} + \delta \omega^{\textrm{nres}}_{10}\\
= \frac{\mu_0 \tilde{\omega}^2_{10} d^2}{8 \pi \hbar} \left[ \frac{1}{z} \sin \left( \frac{2 \tilde{\omega}_{10} z}{c} \right) + \frac{c}{2 \tilde{\omega}_{10} z^2} \cos \left( \frac{2 \tilde{\omega}_{10} z}{c} \right) \right]\\
+\frac{\mu_0 d^2}{8 \pi^2 \hbar} \int\limits^{\infty}_0{\dif \xi \: \frac{\xi^3}{\xi^2 + \tilde{\omega}^2_{10}} \left( \frac{1}{z} + \frac{c}{2 \xi z^2} \right) \me^{-\frac{2 \xi z}{c}}}.
\label{eq:Frequency Shift Perfect Nonreciprocal Medium}
\end{multline}
In the retarded and nonretarded limits the nonresonant part has the asymptotic behavior
\begin{align}
\begin{array}{lll}
&\delta \omega^{\textrm{nres}}_{10} &= \begin{cases} \displaystyle{\frac{d^2 c^2}{16 \pi^2 \epsilon_0 \hbar \tilde{\omega}^2_{10} z^5}}, & \displaystyle{\frac{\tilde{\omega}_{10} z}{c} \gg 1}, \\[2mm] \displaystyle{\frac{d^2}{16 \pi^2 \epsilon_0 \hbar z^3}}, & \displaystyle{\frac{\tilde{\omega}_{10} z}{c} \ll 1}, \end{cases}
\end{array}
\end{align}
which is shown in a double logarithmic plot in Fig.~\ref{fig:PlotLogLog}. The resonant part has the limits
\begin{align}
\begin{array}{lll}
&\delta \omega^{\textrm{res}}_{10} &= \begin{cases} \displaystyle{\frac{\mu_0 \tilde{\omega}^2_{10} d^2}{8 \pi \hbar z} \sin \left( \frac{2 \tilde{\omega}_{10} z}{c} \right)}, & \displaystyle{\frac{\tilde{\omega}_{10} z}{c} \gg 1}, \\[2mm] \displaystyle{\frac{\mu_0 \tilde{\omega}_{10} d^2 c}{16 \pi \hbar z^2}}, & \displaystyle{\frac{\tilde{\omega}_{10} z}{c} \ll 1}. \end{cases}\\[2mm]
\end{array}
\end{align}
By comparing both the decay rates and the resonant frequency shifts in Figs.~\ref{fig:PlotGamma} and \ref{fig:PlotOmega}, a phase shift by $\pi/2$ between the the respective curves of the perfectly conducting mirror and the perfectly reflecting nonreciprocal mirror is apparent, as can also be read off from the first terms in Eqs.~\eqref{eq:Decay Rate Perfectly Reflecting Mirror}, \eqref{eq:Frequency Shift Perfectly Reflecting Mirror}, \eqref{eq:Decay Rate Perfect Nonreciprocal Medium} and \eqref{eq:Frequency Shift Perfect Nonreciprocal Medium}. This is the additional phase shift implied by the reflection of s- into p-polarized waves. The scaling behavior of the decay rates and the resonant frequency shifts in the retarded limit is the same. The decay of the resonant frequency shift in the nonretarded limit is proportional to $z^{-3}$ for the perfectly conducting mirror and $z^{-2}$ for the perfectly reflecting nonreciprocal mirror. As for the nonresonant frequency shift \eqref{eq:Frequency Shift Perfectly Reflecting Mirror} and \eqref{eq:Frequency Shift Perfect Nonreciprocal Medium}, the perfectly reflecting nonreciprocal mirror decays with $z^{-5}$ in contrast to $z^{-4}$ for the perfectly conducting mirror in the retarded limit. The scaling behavior in the nonretarded limit is $z^{-3}$ for both media.\\
The term $z^{-1}$ of the total frequency shift, the sum of the resonant and the nonresonant part, will dominate in the retarded limit both for the perfectly conducting mirror and the perfectly reflecting nonreciprocal mirror. As for the total frequency shift in the nonretarded limit, there is a dominant $z^{-3}$ scaling behavior for both ideal materials.

\subsection{Topological Insulator}
\label{sec:Time-reversal Breaking Topological Insulator}
\begin{figure}[!ht]
\centerline{\includegraphics[width=\columnwidth]{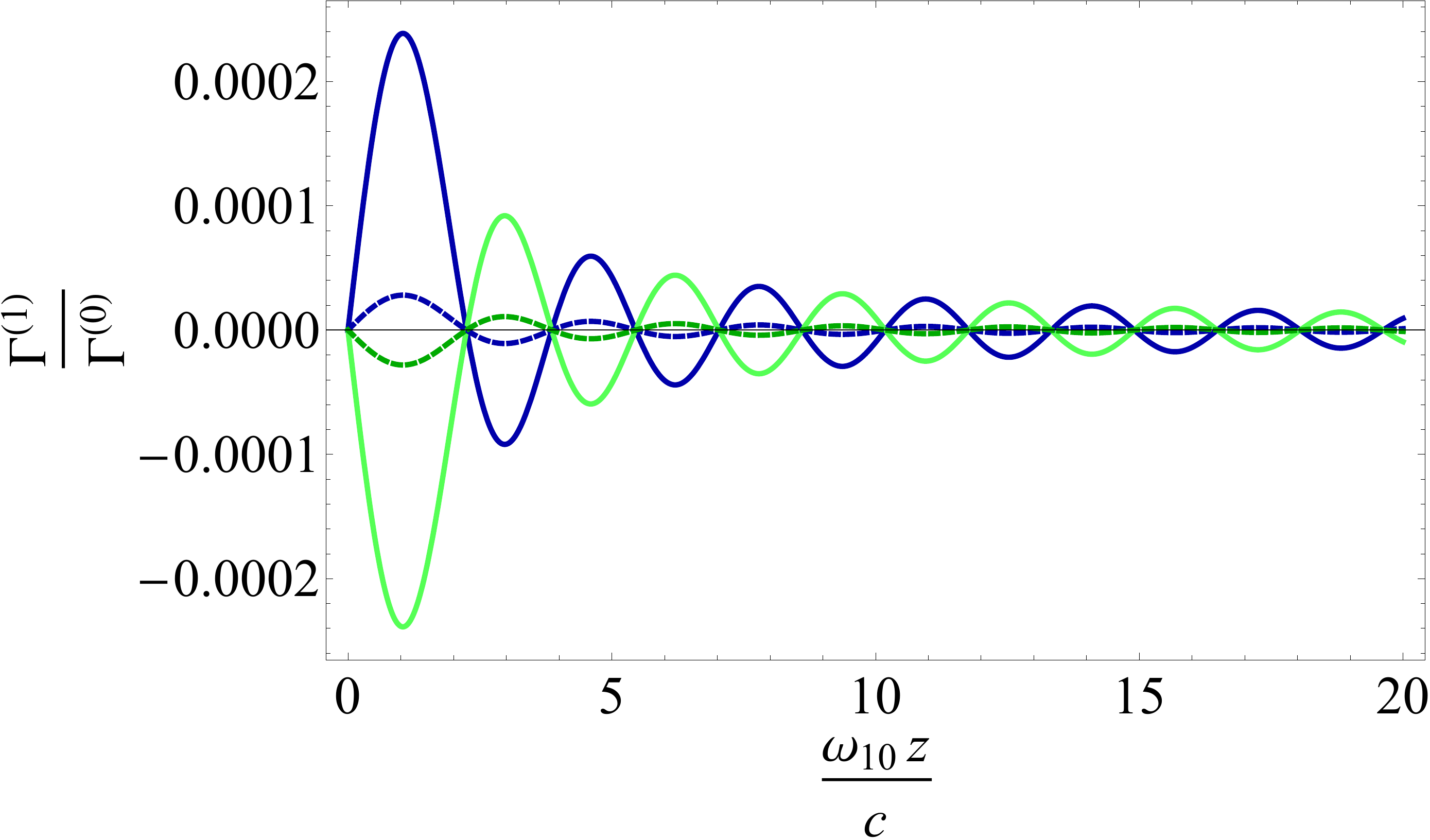}}
\caption{Atomic decay rates $\Gamma^{(1)}$ for a circularly polarized two-level atomic dipole in front of a topological insulator with $\theta = \pi$ \eqref{eq:Free Space Decay Rate} ($\textcolor{LimeGreen}{\hdashrule[0.5ex][x]{0.6cm}{1pt}{}}$) and $\theta = -\pi$ ($\textcolor{Blue}{\hdashrule[0.5ex][x]{0.6cm}{1pt}{}}$) and $\epsilon = \mu = 1$ scaled by the free-space decay rate $\Gamma^{(0)}$. The difference between the decay rate for $\textrm{Bi}_2\textrm{Se}_3$ with axion contribution ($\epsilon = 16, \mu = 1$) and the respective decay rate without axion contribution is depicted for an axion coupling of $\theta = \pi$ ($\textcolor{LimeGreen}{\hdashrule[0.5ex][x]{0.6cm}{1pt}{0.2mm}}$) and $\theta = -\pi$ ($\textcolor{Blue}{\hdashrule[0.5ex][x]{0.6cm}{1pt}{0.2mm}}$).}
\label{fig:PlotGammaTI}
\end{figure}

\begin{figure}[!ht]
\centerline{\includegraphics[width=\columnwidth]{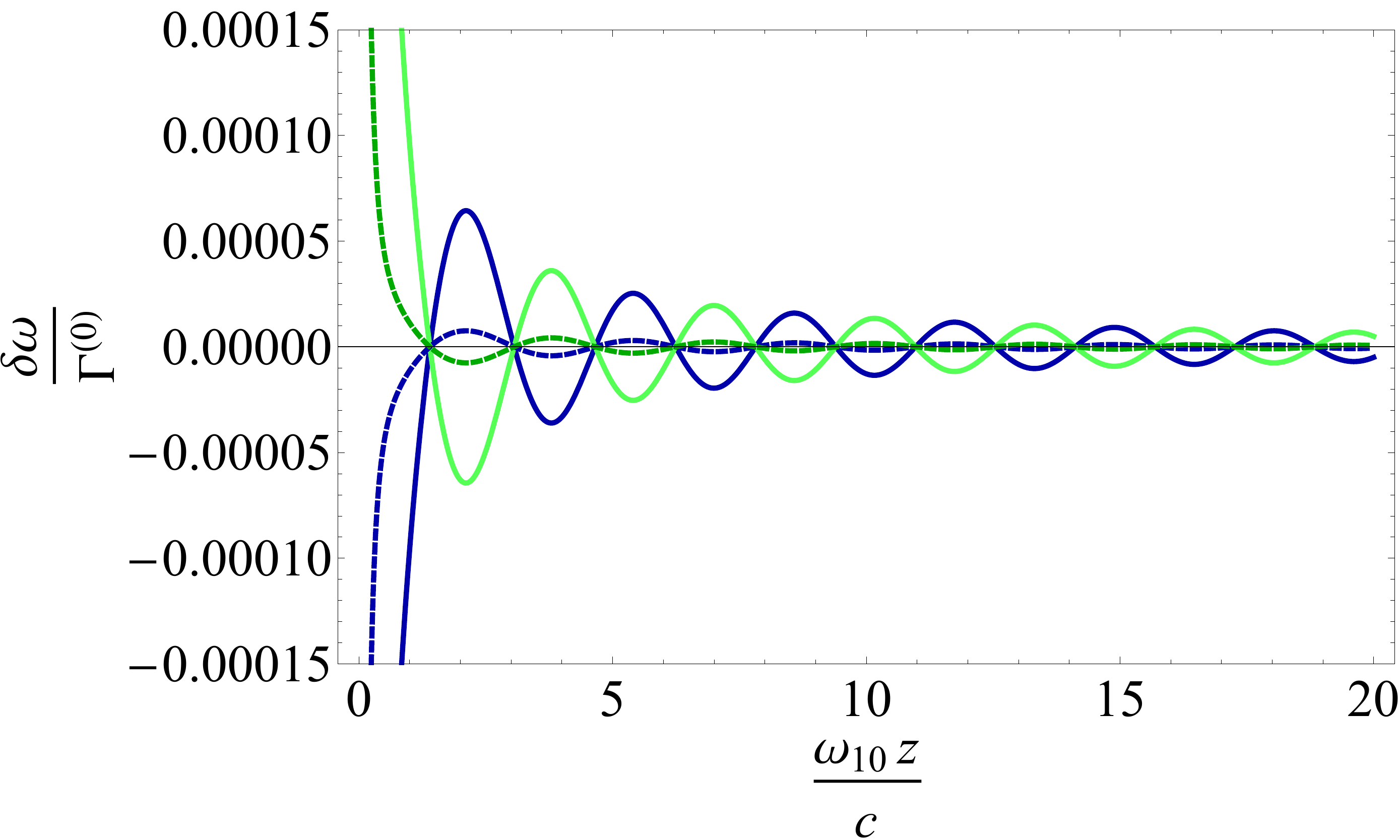}}
\caption{Resonant frequency shift $\delta \omega^{\textrm{res}}$ for a circularly polarized two-level atomic dipole in front of a topological insulator \eqref{eq:Free Space Decay Rate} with $\theta = \pi$ ($\textcolor{LimeGreen}{\hdashrule[0.5ex][x]{0.6cm}{1pt}{}}$) and $\theta = -\pi$ ($\textcolor{Blue}{\hdashrule[0.5ex][x]{0.6cm}{1pt}{}}$) and $\epsilon = \mu = 1$ scaled by the free-space decay rate $\Gamma^{(0)}$. The difference between the resonant frequency shift for $\textrm{Bi}_2\textrm{Se}_3$ with axion contribution ($\epsilon = 16, \mu = 1$) and the respective frequency shift without axion contribution is depicted for an axion coupling of $\theta = \pi$ ($\textcolor{LimeGreen}{\hdashrule[0.5ex][x]{0.6cm}{1pt}{0.2mm}}$) and $\theta = -\pi$ ($\textcolor{Blue}{\hdashrule[0.5ex][x]{0.6cm}{1pt}{0.2mm}}$).}
\label{fig:PlotOmegaTI}
\end{figure}

Ref.~\cite{Crosse:2015} studies the electromagnetic behavior of a topological insulator. Permittivity, permeability and the magneto-electric cross-susceptibilities for this material mentioned in Eq.~\eqref{eq:M Matrix} are assigned according to
\begin{align}
\begin{array}{lll}
&\textrm{\Greektens{$\upepsilon$}} - \textrm{\Greektens{$\upxi$}} \star \textrm{\Greektens{$\upmu$}}^{-1} \star \textrm{\Greektens{$\upzeta$}} \rightarrow \textrm{\Greektens{$\upepsilon$}}, & \textrm{\Greektens{$\upxi$}} \star \textrm{\Greektens{$\upmu$}}^{-1} \rightarrow \displaystyle{\frac{\alpha}{\pi}} \theta \left( \vec{r}, \omega \right)\\[2mm]
&\textrm{\Greektens{$\upmu$}}^{-1} \star \textrm{\Greektens{$\upzeta$}} \rightarrow \displaystyle{\frac{\alpha}{\pi}} \theta \left( \vec{r}, \omega \right), &\textrm{\Greektens{$\upmu$}}^{-1} \rightarrow \textrm{\Greektens{$\upmu$}}^{-1},
\end{array}
\end{align}
so that Eq.~\eqref{eq:Constitutive Relations} takes the form
\begin{alignat}{3}
&\hat{\vec{D}} &=& \epsilon_0 \epsilon \hat{\vec{E}} + \frac{\alpha}{\pi} \frac{\theta \left( \vec{r}, \omega \right)}{\mu_0 c} \hat{\vec{B}} + \hat{\vec{P}}_{\textrm{N}}\\
&\hat{\vec{H}} &=& -\frac{\alpha}{\pi} \frac{\theta \left( \vec{r}, \omega \right)}{\mu_0 c} \hat{\vec{E}} + \frac{1}{\mu_0 \mu} \hat{\vec{B}} - \hat{\vec{M}}_{\textrm{N}}.
\end{alignat}
The reflective coefficients $r_{\sigma,\sigma'}$ mentioned in Eq.~\eqref{eq:Green's Tensor Scattering Part} are given by \cite{Crosse:2015}
\begin{align}
\begin{array}{lll}
&r_{\textrm{s},\textrm{s}} &= \displaystyle{\frac{\left( k^{\perp}_1 - k^{\perp}_2 \right) \left( k^{\perp}_1 \epsilon_2 + k^{\perp}_2 \epsilon_1 \right) - k^{\perp}_1 k^{\perp}_2 \Delta^2}{\left( k^{\perp}_1 + k^{\perp}_2\right) \left( k^{\perp}_1 \epsilon_2 + k^{\perp}_2 \epsilon_1 \right) + k^{\perp}_1 k^{\perp}_2 \Delta^2}}\\[4mm]
&r_{\textrm{p},\textrm{s}} &= \displaystyle{\frac{2 k^{\perp}_1 k^{\perp}_2 n_1 \Delta}{\left( k^{\perp}_1 + k^{\perp}_2 \right) \left( k^{\perp}_1 \epsilon_2 + k^{\perp}_2 \epsilon_1 \right) + k^{\perp}_1 k^{\perp}_2 \Delta^2}}\\[4mm]
&r_{\textrm{p},\textrm{p}} &= \displaystyle{\frac{\left( \epsilon_2 k^{\perp}_1 - \epsilon_1 k^{\perp}_2 \right) \left( k^{\perp}_1 + k^{\perp}_2 \right)+k^{\perp}_1 k^{\perp}_2 \Delta^2}{\left( \epsilon_2 k^{\perp}_1 + \epsilon_1 k^{\perp}_2 \right) \left( k^{\perp}_1 + k^{\perp}_2 \right) + k^{\perp}_1 k^{\perp}_2 \Delta^2}}\\[4mm]
&r_{\textrm{s},\textrm{p}} &= \displaystyle{\frac{2 n_1 k^{\perp}_1 k^{\perp}_2 \Delta}{\left( \epsilon_2 k^{\perp}_1 + \epsilon_1 k^{\perp}_2 \right) \left( k^{\perp}_1 + k^{\perp}_2 \right) + k^{\perp}_1 k^{\perp}_2 \Delta^2}}.
\label{eq:Reflective Coefficients Topological Insulator}
\end{array}
\end{align}
In these equations $k^{\perp}_1$ and $k^{\perp}_2$ refer to the perpendicular part of the wave vector in medium 1 and 2 and $\Delta$ is given by
\begin{equation}
\Delta = \alpha \frac{1}{\pi} \left( \theta_2 - \theta_1 \right),
\end{equation}
where $\alpha$ represents the fine-structure constant and $\theta_1$ and $\theta_2$ are the axion coupling constants in the two media.\\
From now on, we assume the first medium to be vacuum and only the second medium has specific electromagnetic properties $(\epsilon_1 = 1, \epsilon_2 \equiv \epsilon, \theta_1 = 0, \theta_2 \equiv \theta)$. Due to the small value of the fine-structure constant $\alpha$ and the small effect on the reflection coefficients \eqref{eq:Reflective Coefficients Topological Insulator}, we first study a purely axion medium by setting $\epsilon = 1$ and $\mu = 1$. Fig.~\ref{fig:PlotGammaTI} shows the atomic decay rate for $\theta = \pi$ and $\theta = -\pi$. Fig.~\ref{fig:PlotOmegaTI} depicts the respective resonant part of the frequency shift. The results for the decay rate and the resonant frequency shift resemble the respective curves of the perfectly reflecting nonreciprocal mirror in Figs.~\ref{fig:PlotGamma} and \ref{fig:PlotOmega}, but are scaled by $\Delta/2$. This ratio can be easily read off in the retarded and nonretarded limits.
The reflective coefficients \eqref{eq:Reflective Coefficients Topological Insulator} in the retarded limit read
\begin{align}
\begin{array}{lll}
&r^{\textrm{ret}}_{\textrm{s},\textrm{s}} &= \displaystyle{\frac{\left( 1 - \epsilon \right) - \Delta^2}{\left( 1 + n \right)^2 + \Delta^2}}\\[4mm]
&r^{\textrm{ret}}_{\textrm{p},\textrm{s}} &= \displaystyle{\frac{-2 \Delta}{\left( 1 + n \right)^2 + \Delta^2}}\\[4mm]
&r^{\textrm{ret}}_{\textrm{p},\textrm{p}} &= \displaystyle{\frac{-\left( 1- \epsilon \right) + \Delta^2}{\left( 1 + n \right)^2 + \Delta^2}} = - r^{\textrm{ret}}_{\textrm{s},\textrm{s}}\\[4mm]
&r^{\textrm{ret}}_{\textrm{s},\textrm{p}} &= \displaystyle{\frac{-2 \Delta}{\left( 1 + n \right)^2 + \Delta^2}} = r^{\textrm{ret}}_{\textrm{p},\textrm{s}}
\label{eq:Reflective Coefficients Topological Insulator Retarded}
\end{array}
\end{align}
with the refractive index $n = \sqrt{\epsilon}$. The decay rate and resonant frequency shift with purely axion contribution in the retarded limit are given by
\begin{align}
\begin{array}{lll}
&\Gamma^{(1) \textrm{ret}}_{10} \left( \epsilon=1 \right) &= \displaystyle{\frac{\mu_0 \tilde{\omega}^2_{10} d^2}{4 \pi \hbar z}} \cos \left( \frac{2 \tilde{\omega}_{10} z}{c} \right) \frac{\Delta}{2}\\[4mm]
&\delta \omega^{\textrm{res,ret}}_{10} \left( \epsilon=1 \right) &= \displaystyle{\frac{\mu_0 \tilde{\omega}^2_{10} d^2}{8 \pi \hbar z}} \sin \left( \frac{2 \tilde{\omega}_{10} z}{c} \right) \frac{\Delta}{2}.
\label{eq:Decay Rate Frequency Shift Retarded}
\end{array}
\end{align}
The same procedure is carried out in the nonretarded limit and the respective reflective coefficients \eqref{eq:Reflective Coefficients Topological Insulator} are
\begin{align}
\begin{array}{lll}
&r^{\textrm{nonret}}_{\textrm{s},\textrm{s}} &= \displaystyle{\frac{- \Delta^2}{2 \left( \epsilon + 1 \right) + \Delta^2}}\\[4mm]
&r^{\textrm{nonret}}_{\textrm{p},\textrm{s}} &= \displaystyle{\frac{-2 \Delta}{2 \left( \epsilon + 1 \right) + \Delta^2}}\\[4mm]
&r^{\textrm{nonret}}_{\textrm{p},\textrm{p}} &= \displaystyle{\frac{2 \left( \epsilon - 1 \right) + \Delta^2}{2 \left( \epsilon + 1 \right) + \Delta^2}}\\[4mm]
&r^{\textrm{nonret}}_{\textrm{s},\textrm{p}} &= \displaystyle{\frac{-2 \Delta}{2 \left( \epsilon + 1 \right) + \Delta^2}} = r^{\textrm{nonret}}_{\textrm{p},\textrm{s}}.
\label{eq:Reflective Coefficients Topological Insulator Nonretarded}
\end{array}
\end{align}
The respective decay rate and resonant frequency shift for the purely axion contribution in the nonretarded limit read
\begin{align}
\begin{array}{lll}
&\Gamma^{(1) \textrm{nonret}}_{10} \left( \epsilon=1 \right) &= -\displaystyle{\frac{\mu_0 \tilde{\omega}_{10} d^2 c}{8 \pi \hbar z^2}} \frac{\Delta}{2}\\
&\delta \omega^{\textrm{res,nonret}}_{10} \left( \epsilon=1 \right) &= \displaystyle{\frac{\mu_0 \tilde{\omega}_{10} d^2 c}{16 \pi \hbar z^2}} \frac{\Delta}{2}.
\label{eq:Decay Rate Frequency Shift Nonretarded}
\end{array}
\end{align}
Next, we look at general material properties similar to $\textrm{Bi}_2\textrm{Se}_3$, where we take $\epsilon = 16$ and $\mu = 1$ \cite{Crosse:2015}. We compare the case with axion coupling of $\theta = \pi$ and without axion coupling $\theta = 0$. Because of the small value of $\alpha$, the reflective coefficients $r_{\textrm{p}, \textrm{s}}$ and $r_{\textrm{s}, \textrm{p}}$ do not have a big impact on the decay rate and the frequency shift. The decay rate and resonant frequency shift in the retarded limit are calculated by inserting the reflective coefficients \eqref{eq:Reflective Coefficients Topological Insulator Retarded} into Eqs.~\eqref{eq:Decay Rate} and \eqref{eq:Frequency Shift Resonant Part}
\begin{align}
\begin{array}{lll}
&\Gamma^{(1) \textrm{ret}}_{10} &= \displaystyle{\frac{\mu_0 \tilde{\omega}^2_{10} d^2}{4 \pi \hbar}} \left[ -\displaystyle{\frac{1}{z}} \sin \left( \frac{2 \tilde{\omega}_{10} z}{c} \right) r^{\textrm{ret}}_{\textrm{p},\textrm{p}} \right.\\[4mm]
& &\left. - \displaystyle{\frac{1}{z}} \cos \left( \frac{2 \tilde{\omega}_{10} z}{c} \right) r^{\textrm{ret}}_{\textrm{s},\textrm{p}} \right]\\[4mm]
&\delta \omega^{\textrm{res,ret}}_{10} &= \displaystyle{\frac{\mu_0 \tilde{\omega}^2_{10} d^2}{8 \pi \hbar}} \left[ \frac{1}{z} \cos \left( \frac{2 \tilde{\omega}_{10} z}{c} \right) r^{\textrm{ret}}_{\textrm{p},\textrm{p}} \right.\\[4mm]
& & \left. - \displaystyle{\frac{1}{z}} \sin \left( \frac{2 \tilde{\omega}_{10} z}{c} \right) r^{\textrm{ret}}_{\textrm{s},\textrm{p}} \right].
\end{array}
\end{align}
The difference in the decay rate and the resonant frequency shift between the cases with and without axion coupling in the retarded limit and for $\Delta \ll 1$ yields
\begin{align}
\begin{array}{lll}
&\Delta \Gamma^{(1) \textrm{ret}}_{10} &\equiv \Gamma^{(1) \textrm{ret}}_{10} - \Gamma^{(1) \textrm{ret}}_{10} \left( \theta = 0 \right)\\[2mm]
& &= \displaystyle{\frac{\mu_0 \tilde{\omega}^2_{10} d^2}{4 \pi \hbar z}} \cos \left( \frac{2 \tilde{\omega}_{10} z}{c} \right) \frac{2 \Delta}{\left( 1+n \right)^2}\\[4mm]
&\Delta \delta \omega^{\textrm{res,ret}}_{10} &\equiv \delta \omega^{\textrm{res,ret}}_{10} - \delta \omega ^{\textrm{res,ret}}_{10} \left( \theta = 0 \right)\\[2mm]
& &= \displaystyle{\frac{\mu_0 \tilde{\omega}^2_{10} d^2}{8 \pi \hbar z}} \sin \left( \frac{2 \tilde{\omega}_{10} z}{c} \right) \frac{2 \Delta}{\left( 1+n \right)^2}.
\label{eq:Decay Rate Frequency Shift Retarded Difference}
\end{array}
\end{align}
In the limit of $\Delta \ll 1$, the scaling factor for Eq.~\eqref{eq:Decay Rate Frequency Shift Retarded Difference} for $\epsilon=16$ with respect to the purely axion material in the retarded limit \eqref{eq:Decay Rate Frequency Shift Retarded} is $4/25$.\\
The decay rate and resonant frequency shift in the nonretarded limit are obtained by inserting Eq.~\eqref{eq:Reflective Coefficients Topological Insulator Nonretarded} into Eqs.~\eqref{eq:Decay Rate} and \eqref{eq:Frequency Shift Resonant Part}
\begin{align}
\begin{array}{lll}
&\Gamma^{(1) \textrm{nonret}}_{10} &= \displaystyle{\frac{\mu_0 \tilde{\omega}^2_{10} d^2}{4 \pi \hbar}} \left[ \displaystyle{\frac{c}{2 \tilde{\omega}_{10} z^2}} r^{\textrm{nonret}}_{\textrm{s},\textrm{p}} +\displaystyle{\frac{c^2}{4 \tilde{\omega}^2_{10} z^3}} r^{\textrm{nonret}}_{\textrm{p},\textrm{p}} \right]\\[4mm]
&\delta \omega^{\textrm{res,nonret}}_{10} &= \displaystyle{\frac{\mu_0 \tilde{\omega}^2_{10} d^2}{8 \pi \hbar}} \left[ -\displaystyle{\frac{c}{2 \tilde{\omega}_{10} z^2}} r^{\textrm{nonret}}_{\textrm{s},\textrm{p}} -\displaystyle{\frac{c^2}{4 \tilde{\omega}^2_{10} z^3}} r^{\textrm{nonret}}_{\textrm{p},\textrm{p}} \right].
\end{array}
\end{align}
The differential effects of the axion coupling on the decay rate and the resonant frequency shift for the nonretarded limit are given by
\begin{align}
\begin{array}{lll}
&\Delta \Gamma^{(1) \textrm{nonret}}_{10} &\equiv \Gamma^{(1) \textrm{nonret}}_{10} - \Gamma^{(1) \textrm{nonret}}_{10} \left( \theta = 0 \right)\\[2mm]
& &= -\displaystyle{\frac{\mu_0 \tilde{\omega}_{10} d^2 c}{8 \pi \hbar z^2}} \frac{\Delta}{\epsilon + 1}\\[4mm]
&\Delta \delta \omega^{\textrm{res,nonret}}_{10} &\equiv \delta \omega^{\textrm{res,nonret}}_{10} - \delta \omega^{\textrm{res,nonret}}_{10} \left( \theta = 0 \right)\\[2mm]
& &= \displaystyle{\frac{\mu_0 \tilde{\omega}_{10} d^2 c}{16 \pi \hbar z^2} \frac{\Delta}{\epsilon + 1}}.
\label{eq:Decay Rate Frequency Shift Nonretarded Difference}
\end{array}
\end{align}
The respective scaling factor of Eq.~\eqref{eq:Decay Rate Frequency Shift Nonretarded Difference} with respect to Eq.~\eqref{eq:Decay Rate Frequency Shift Nonretarded} in the nonretarded limit for $\epsilon = 16$ is $2/17$. The difference in the atomic decay rate and the resonant frequency shift between these two cases follows the same form of the purely axion atomic decay rate and frequency shift and can be compared to that. The scaling factors are gauged in the retarded and nonretarded limit, cf. Figs.~\ref{fig:PlotGammaTI} and \ref{fig:PlotOmegaTI}.\\
The nonresonant frequency shift \eqref{eq:Frequency Shift Nonresonant Part} for the topological insulator contains frequency-dependent permeability and permittivity $\epsilon \left( \mi \xi \right)$ and $\mu \left( \mi \xi \right)$. Without knowing the exact behavior of these quantities we can only approximate the nonresonant frequency shift in the retarded and nonretarded limits. Since the resonant frequency shift always dominates in the retarded limit, we restrict ourselves to gauge the nonretarded limit. For a purely nonreciprocal medium with $\epsilon = 1$, we obtain
\begin{equation}
\delta \omega^{\textrm{nres},\textrm{nonret}}_{10} \left( \epsilon = 1 \right) = \frac{d^2}{16 \pi^2 \epsilon_0 \hbar z^3} \frac{\Delta}{2}.
\end{equation}
The result for a general medium reads
\begin{multline}
\delta \omega^{\textrm{nres},\textrm{nonret}}_{10} = \frac{\mu_0 d^2}{8 \pi^2 \hbar} \int\limits^{\infty}_0 \dif \xi \frac{\xi^2}{\xi^2 + \tilde{\omega}^2_{10}} \me^{-\frac{2 \xi z}{c}}\\
\left\{ -\frac{\xi}{z} r^{\textrm{nonret}}_{\textrm{s},\textrm{p}} - \frac{c}{2 z^2} r^{\textrm{nonret}}_{\textrm{s},\textrm{p}} + \frac{c^2 \tilde{\omega}_{10}}{4 \xi^2 z^3} r^{\textrm{nonret}}_{\textrm{p},\textrm{p}} \right\}.
\end{multline}
Because of the strong effect of $\epsilon$ compared to $\Delta$ the terms with $r^{\textrm{nonret}}_{\textrm{p}, \textrm{p}}$ \eqref{eq:Reflective Coefficients Topological Insulator Nonretarded} do not have to be considered for the difference between the topological insulator with and without axion coupling. Only $r^{\textrm{nonret}}_{\textrm{s},\textrm{p}}$ remains and is inserted into Eq.~\eqref{eq:Frequency Shift Nonresonant Part}. For $\xi \rightarrow \infty$, $\epsilon \left( \mi \xi \right) \rightarrow 1$. After performing the $\xi$-integral, we obtain the final result for the difference of the nonresonant frequency shift of the topological insulator in the nonretarded limit
\begin{align}
\begin{array}{lll}
&\Delta \delta \omega^{\textrm{nres},\textrm{nonret}}_{10} &\equiv \delta \omega^{\textrm{nres},\textrm{nonret}}_{10} - \delta \omega^{\textrm{nres},\textrm{nonret}}_{10} \left( \theta = 0 \right)\\[2mm]
& &= \displaystyle{\frac{d^2}{16 \pi^2 \epsilon_0 \hbar z^3}} \frac{\Delta}{\epsilon + 1}.
\end{array}
\end{align}
The total frequency shift of the resonant and nonresonant parts of the topological insulator scales with $z^{-1}$ in the nonretarded limit. An experimental distinction from another material is difficult, cf. Sec. \ref{sec:Perfect Nonreciprocal Medium}.\\
In case of an extremely large axion coupling, the reflective coefficients \eqref{eq:Reflective Coefficients Topological Insulator} reduce to the values $r_{\textrm{s,s}}=-1$ and $r_{\textrm{p,p}}=1$. Both the decay rates, the resonant frequency shift and the nonresonant decay shift approximate the results of the perfectly conducting mirror, cf. Figs.~\ref{fig:PlotGamma} and \ref{fig:PlotOmega}.\\
Note that for each of the interacting time-reversal symmetry breaking subsystems, atom and medium, there are two possible choices regarding their internal sense of time. For the atom, they correspond to clockwise versus counterclockwise circular dipole transitions and can be related to one another via $\vec{d} \rightarrow \vec{d}^*$. For the medium, the two possible internal senses of time are related via $\Delta \rightarrow - \Delta$ or $r_{\textrm{s,p}}$, $r_{\textrm{p,s}}$ $\rightarrow$ $-r_{\textrm{s,p}},-r_{\textrm{p,s}}$. We thus have four possible combinations of t-odd atoms interacting with nonreciprocal media. The other three possible combinations can be obtained from the particular choice considered here by changing the internal arrow of time in atom, medium or both, where each such change reverses the signs of frequency shift and body-assisted decay rate.\\
Due to the internal connection between the frequency shift and the Casmir--Polder force, one can also switch from an attractive to a repulsive force between atom and medium.
	
\section{Summary}
We have applied macroscopic QED to derive expressions for the Casimir--Polder frequency shift and spontaneous decay rate for nonreciprocal media, that violate Lorentz reciprocity principle and therefore break time-reversal symmetry. Consequently, real and imaginary parts of the Green's tensor for nonreciprocal media have to be redefined by using the adjoint tensor instead of the complex conjugate one.\\
Based on the interaction Hamiltonian between the atom, the field and the nonreciprocal medium, an expression for the electric field has been obtained in two alternative ways. First, noise currents can be quantized directly yielding one set of field operators for the combined electric and magnetic fields. According to the second approach the noise currents can be divided into contributions for the polarization and the magnetization giving rise to cross-correlations between electric and magnetic fields. The result for the electric field has enabled us to study the internal atomic dynamics. By making use of the redefined real and imaginary parts of a tensor, we obtain general expressions for the atomic decay rate and the frequency shift, which can be split into a resonant and a nonresonant contribution, representing generalizations for nonreciprocal media.\\
As an example, we have investigated the decay rate and frequency shift for a two-level atom with circularly polarized dipole moments in order to be able to detect the broken time-reversal symmetry. First, a perfectly conducting mirror has been compared to a perfectly reflecting nonreciprocal mirror yielding different polynomial scaling behaviors. Whereas the nonresonant frequency shift of the perfectly conducting mirror decays with $z^{-4}$ in the retarded limit, it scales with $z^{-5}$ in case of the perfectly reflecting nonreciprocal mirror. In the nonretarded limit both scale with $z^{-3}$. As for the resonant frequency shift, there is a $z^{-1}$ behavior for both materials in the retarded limit and in the nonretarded limit they differ again, the perfectly conducting mirror scales with $z^{-3}$, the perfectly reflecting nonreciprocal mirror with $z^{-2}$\\
Second, we have investigated a time-reversal-symmetry-broken topological insulator, whose electromagnetic properties are described by an axion coupling and whose reflective coefficients depend on the wave vector. Due to the small impact of the axion part, we have restricted ourselves to a medium of pure axion behavior by setting $\epsilon = 1$ and compared this to the difference quantities between included axion coupling and without axion coupling for a material similar to $\textrm{Bi}_2\textrm{Se}_3$. We find a qualitatively similar behavior and determine scaling factors between the two cases in the retarded and nonretarded limits. Finally, we can switch the sign of the decay rate and the frequency shift of the topological insulator both by reversing the direction of the oscillating dipole moments and by changing the sign of the axion coupling. This opens the door for switching between attractive and repulsive Casimir--Polder forces.
	
\section{acknowledgments}
We acknowledge helpful discussions with Tobias Br\"unner, Robert Bennett, Diego Dalvit and Homer Reid. This work was supported by the German Research Foundation (DFG, Grants BU 1803/3-1 and GRK 2079/1). S.Y.B is grateful for support by the Freiburg Institute of Advanced Studies.



\begin{thebibliography}{42}%
\makeatletter
\providecommand \@ifxundefined [1]{%
 \@ifx{#1\undefined}
}%
\providecommand \@ifnum [1]{%
 \ifnum #1\expandafter \@firstoftwo
 \else \expandafter \@secondoftwo
 \fi
}%
\providecommand \@ifx [1]{%
 \ifx #1\expandafter \@firstoftwo
 \else \expandafter \@secondoftwo
 \fi
}%
\providecommand \natexlab [1]{#1}%
\providecommand \enquote  [1]{``#1''}%
\providecommand \bibnamefont  [1]{#1}%
\providecommand \bibfnamefont [1]{#1}%
\providecommand \citenamefont [1]{#1}%
\providecommand \href@noop [0]{\@secondoftwo}%
\providecommand \href [0]{\begingroup \@sanitize@url \@href}%
\providecommand \@href[1]{\@@startlink{#1}\@@href}%
\providecommand \@@href[1]{\endgroup#1\@@endlink}%
\providecommand \@sanitize@url [0]{\catcode `\\12\catcode `\$12\catcode
  `\&12\catcode `\#12\catcode `\^12\catcode `\_12\catcode `\%12\relax}%
\providecommand \@@startlink[1]{}%
\providecommand \@@endlink[0]{}%
\providecommand \url  [0]{\begingroup\@sanitize@url \@url }%
\providecommand \@url [1]{\endgroup\@href {#1}{\urlprefix }}%
\providecommand \urlprefix  [0]{URL }%
\providecommand \Eprint [0]{\href }%
\providecommand \doibase [0]{http://dx.doi.org/}%
\providecommand \selectlanguage [0]{\@gobble}%
\providecommand \bibinfo  [0]{\@secondoftwo}%
\providecommand \bibfield  [0]{\@secondoftwo}%
\providecommand \translation [1]{[#1]}%
\providecommand \BibitemOpen [0]{}%
\providecommand \bibitemStop [0]{}%
\providecommand \bibitemNoStop [0]{.\EOS\space}%
\providecommand \EOS [0]{\spacefactor3000\relax}%
\providecommand \BibitemShut  [1]{\csname bibitem#1\endcsname}%
\let\auto@bib@innerbib\@empty
\bibitem [{\citenamefont {Casimir}\ and\ \citenamefont
  {Polder}(1948)}]{Casimir_Polder:1948}%
  \BibitemOpen
  \bibfield  {author} {\bibinfo {author} {\bibfnamefont {H.}~\bibnamefont
  {Casimir}}\ and\ \bibinfo {author} {\bibfnamefont {D.}~\bibnamefont
  {Polder}},\ }\href {http://journals.aps.org/pr/pdf/10.1103/PhysRev.73.360}
  {\bibfield  {journal} {\bibinfo  {journal} {Physical Review}\ }\textbf
  {\bibinfo {volume} {73}},\ \bibinfo {pages} {360} (\bibinfo {year}
  {1948})}\BibitemShut {NoStop}%
\bibitem [{\citenamefont {Casimir}(1948)}]{Casimir:1948}%
  \BibitemOpen
  \bibfield  {author} {\bibinfo {author} {\bibfnamefont {H.~B.~G.}\
  \bibnamefont {Casimir}},\ }\href
  {http://www.mit.edu/~kardar/research/seminars/Casimir/Casimir1948.pdf}
  {\bibfield  {journal} {\bibinfo  {journal} {Proc. K. Ned. Acad Wet.}\
  }\textbf {\bibinfo {volume} {51}},\ \bibinfo {pages} {793} (\bibinfo {year}
  {1948})}\BibitemShut {NoStop}%
\bibitem [{\citenamefont {Buhmann}(2012{\natexlab{a}})}]{Buhmann_Book_1}%
  \BibitemOpen
  \bibfield  {author} {\bibinfo {author} {\bibfnamefont {S.~Y.}\ \bibnamefont
  {Buhmann}},\ }\href@noop {} {\emph {\bibinfo {title} {Dispersion Forces I -
  Macroscopic Quantum Electrodynamics and Ground-State Casimir, Casimir--Polder
  and van der Waals Forces}}}\ (\bibinfo  {publisher} {Springer, Berlin
  Heidelberg},\ \bibinfo {year} {2012})\BibitemShut {NoStop}%
\bibitem [{\citenamefont {Buhmann}(2012{\natexlab{b}})}]{Buhmann_Book_2}%
  \BibitemOpen
  \bibfield  {author} {\bibinfo {author} {\bibfnamefont {S.~Y.}\ \bibnamefont
  {Buhmann}},\ }\href@noop {} {\emph {\bibinfo {title} {Dispersion Forces II -
  Many-Body Effects, Excited Atoms, Finite Temperature and Quantum Friction}}}\
  (\bibinfo  {publisher} {Springer, Berlin Heidelberg},\ \bibinfo {year}
  {2012})\BibitemShut {NoStop}%
\bibitem [{\citenamefont {Dung}\ \emph {et~al.}(1998)\citenamefont {Dung},
  \citenamefont {Kn{\"o}ll},\ and\ \citenamefont {Welsch}}]{Dung:1998}%
  \BibitemOpen
  \bibfield  {author} {\bibinfo {author} {\bibfnamefont {H.~T.}\ \bibnamefont
  {Dung}}, \bibinfo {author} {\bibfnamefont {L.}~\bibnamefont {Kn{\"o}ll}}, \
  and\ \bibinfo {author} {\bibfnamefont {D.-G.}\ \bibnamefont {Welsch}},\
  }\href {http://journals.aps.org/pra/pdf/10.1103/PhysRevA.57.3931} {\bibfield
  {journal} {\bibinfo  {journal} {Physical Review A}\ }\textbf {\bibinfo
  {volume} {57}},\ \bibinfo {pages} {3931} (\bibinfo {year}
  {1998})}\BibitemShut {NoStop}%
\bibitem [{\citenamefont {Buhmann}\ \emph {et~al.}(2004)\citenamefont
  {Buhmann}, \citenamefont {Kn{\"o}ll}, \citenamefont {Welsch},\ and\
  \citenamefont {Dung}}]{Buhmann:2004}%
  \BibitemOpen
  \bibfield  {author} {\bibinfo {author} {\bibfnamefont {S.~Y.}\ \bibnamefont
  {Buhmann}}, \bibinfo {author} {\bibfnamefont {L.}~\bibnamefont {Kn{\"o}ll}},
  \bibinfo {author} {\bibfnamefont {D.-G.}\ \bibnamefont {Welsch}}, \ and\
  \bibinfo {author} {\bibfnamefont {H.~T.}\ \bibnamefont {Dung}},\ }\href
  {http://journals.aps.org/pra/abstract/10.1103/PhysRevA.70.052117} {\bibfield
  {journal} {\bibinfo  {journal} {Physical Review A}\ }\textbf {\bibinfo
  {volume} {70}},\ \bibinfo {pages} {052117} (\bibinfo {year}
  {2004})}\BibitemShut {NoStop}%
\bibitem [{\citenamefont {Buhmann}\ and\ \citenamefont
  {Welsch}(2007)}]{Buhmann:2007}%
  \BibitemOpen
  \bibfield  {author} {\bibinfo {author} {\bibfnamefont {S.~Y.}\ \bibnamefont
  {Buhmann}}\ and\ \bibinfo {author} {\bibfnamefont {D.-G.}\ \bibnamefont
  {Welsch}},\ }\href {http://arxiv.org/pdf/quant-ph/0608118.pdf} {\bibfield
  {journal} {\bibinfo  {journal} {Progress in quantum electronics}\ }\textbf
  {\bibinfo {volume} {31}},\ \bibinfo {pages} {51} (\bibinfo {year}
  {2007})}\BibitemShut {NoStop}%
\bibitem [{\citenamefont {Scheel}\ and\ \citenamefont
  {Buhmann}(2008)}]{Scheel:2008}%
  \BibitemOpen
  \bibfield  {author} {\bibinfo {author} {\bibfnamefont {S.}~\bibnamefont
  {Scheel}}\ and\ \bibinfo {author} {\bibfnamefont {S.~Y.}\ \bibnamefont
  {Buhmann}},\ }\href
  {http://www.physics.sk/aps/pubs/2008/aps-08-05/aps-08-05.pdf} {\bibfield
  {journal} {\bibinfo  {journal} {Acta Phys. Slovaca}\ }\textbf {\bibinfo
  {volume} {58}},\ \bibinfo {pages} {675} (\bibinfo {year} {2008})}\BibitemShut
  {NoStop}%
\bibitem [{\citenamefont {Skagerstam}\ \emph {et~al.}(2009)\citenamefont
  {Skagerstam}, \citenamefont {Rekdal},\ and\ \citenamefont
  {Vaskinn}}]{Skagerstam:2009}%
  \BibitemOpen
  \bibfield  {author} {\bibinfo {author} {\bibfnamefont {B.-S.}\ \bibnamefont
  {Skagerstam}}, \bibinfo {author} {\bibfnamefont {P.~K.}\ \bibnamefont
  {Rekdal}}, \ and\ \bibinfo {author} {\bibfnamefont {A.~H.}\ \bibnamefont
  {Vaskinn}},\ }\href
  {http://journals.aps.org/pra/pdf/10.1103/PhysRevA.80.022902} {\bibfield
  {journal} {\bibinfo  {journal} {Physical Review A}\ }\textbf {\bibinfo
  {volume} {80}},\ \bibinfo {pages} {022902} (\bibinfo {year}
  {2009})}\BibitemShut {NoStop}%
\bibitem [{\citenamefont {Chew}(1995)}]{Chew_Book}%
  \BibitemOpen
  \bibfield  {author} {\bibinfo {author} {\bibfnamefont {W.~C.}\ \bibnamefont
  {Chew}},\ }\href@noop {} {\emph {\bibinfo {title} {Waves and Fields in
  Inhomogeneous Media}}}\ (\bibinfo  {publisher} {IEEE PRESS Series on
  Electromagnetic Waves},\ \bibinfo {year} {1995})\BibitemShut {NoStop}%
\bibitem [{\citenamefont {Kn{\"o}ll}\ \emph {et~al.}(2001)\citenamefont
  {Kn{\"o}ll}, \citenamefont {Scheel},\ and\ \citenamefont
  {Welsch}}]{Knoll_Book}%
  \BibitemOpen
  \bibfield  {author} {\bibinfo {author} {\bibfnamefont {L.}~\bibnamefont
  {Kn{\"o}ll}}, \bibinfo {author} {\bibfnamefont {S.}~\bibnamefont {Scheel}}, \
  and\ \bibinfo {author} {\bibfnamefont {D.~G.}\ \bibnamefont {Welsch}},\
  }\href@noop {} {\emph {\bibinfo {title} {Coherence and Statistics of Photons
  and Atoms}}},\ edited by\ \bibinfo {editor} {\bibfnamefont {J.}~\bibnamefont
  {Perina}}\ (\bibinfo  {publisher} {Wiley, New York},\ \bibinfo {year}
  {2001})\BibitemShut {NoStop}%
\bibitem [{\citenamefont {Eberlein}\ and\ \citenamefont
  {Zietal}(2012)}]{Eberlein:2012}%
  \BibitemOpen
  \bibfield  {author} {\bibinfo {author} {\bibfnamefont {C.}~\bibnamefont
  {Eberlein}}\ and\ \bibinfo {author} {\bibfnamefont {R.}~\bibnamefont
  {Zietal}},\ }\href
  {http://journals.aps.org/pra/pdf/10.1103/PhysRevA.86.062507} {\bibfield
  {journal} {\bibinfo  {journal} {Physical Review A}\ }\textbf {\bibinfo
  {volume} {86}},\ \bibinfo {pages} {062507} (\bibinfo {year}
  {2012})}\BibitemShut {NoStop}%
\bibitem [{\citenamefont {Xu}\ \emph {et~al.}(2014)\citenamefont {Xu},
  \citenamefont {Alamri}, \citenamefont {Yang}, \citenamefont {Zhu},\ and\
  \citenamefont {Zubairy}}]{Xu:2014}%
  \BibitemOpen
  \bibfield  {author} {\bibinfo {author} {\bibfnamefont {J.}~\bibnamefont
  {Xu}}, \bibinfo {author} {\bibfnamefont {M.}~\bibnamefont {Alamri}}, \bibinfo
  {author} {\bibfnamefont {Y.}~\bibnamefont {Yang}}, \bibinfo {author}
  {\bibfnamefont {S.-Y.}\ \bibnamefont {Zhu}}, \ and\ \bibinfo {author}
  {\bibfnamefont {M.~S.}\ \bibnamefont {Zubairy}},\ }\href
  {http://journals.aps.org/pra/pdf/10.1103/PhysRevA.89.053831} {\bibfield
  {journal} {\bibinfo  {journal} {Physical Review A}\ }\textbf {\bibinfo
  {volume} {89}},\ \bibinfo {pages} {053831} (\bibinfo {year}
  {2014})}\BibitemShut {NoStop}%
\bibitem [{\citenamefont {Henkel}\ and\ \citenamefont
  {Joulain}(2005)}]{Henkel:2005}%
  \BibitemOpen
  \bibfield  {author} {\bibinfo {author} {\bibfnamefont {C.}~\bibnamefont
  {Henkel}}\ and\ \bibinfo {author} {\bibfnamefont {K.}~\bibnamefont
  {Joulain}},\ }\href {http://arxiv.org/pdf/quant-ph/0407153.pdf} {\bibfield
  {journal} {\bibinfo  {journal} {EPL (Europhysics Letters)}\ }\textbf
  {\bibinfo {volume} {72}},\ \bibinfo {pages} {929} (\bibinfo {year}
  {2005})}\BibitemShut {NoStop}%
\bibitem [{\citenamefont {Crosse}\ \emph {et~al.}(2010)\citenamefont {Crosse},
  \citenamefont {Ellingsen}, \citenamefont {Clements}, \citenamefont
  {Buhmann},\ and\ \citenamefont {Scheel}}]{Crosse:2010}%
  \BibitemOpen
  \bibfield  {author} {\bibinfo {author} {\bibfnamefont {J.~A.}\ \bibnamefont
  {Crosse}}, \bibinfo {author} {\bibfnamefont {S.~A.}\ \bibnamefont
  {Ellingsen}}, \bibinfo {author} {\bibfnamefont {K.}~\bibnamefont {Clements}},
  \bibinfo {author} {\bibfnamefont {S.~Y.}\ \bibnamefont {Buhmann}}, \ and\
  \bibinfo {author} {\bibfnamefont {S.}~\bibnamefont {Scheel}},\ }\href
  {http://journals.aps.org/pra/pdf/10.1103/PhysRevA.82.010901} {\bibfield
  {journal} {\bibinfo  {journal} {Physical Review A}\ }\textbf {\bibinfo
  {volume} {82}},\ \bibinfo {pages} {010901} (\bibinfo {year}
  {2010})}\BibitemShut {NoStop}%
\bibitem [{\citenamefont {Onsager}(1931)}]{Onsager:1931}%
  \BibitemOpen
  \bibfield  {author} {\bibinfo {author} {\bibfnamefont {L.}~\bibnamefont
  {Onsager}},\ }\href {http://journals.aps.org/pr/pdf/10.1103/PhysRev.37.405}
  {\bibfield  {journal} {\bibinfo  {journal} {Physical Review}\ }\textbf
  {\bibinfo {volume} {37}},\ \bibinfo {pages} {405} (\bibinfo {year}
  {1931})}\BibitemShut {NoStop}%
\bibitem [{\citenamefont {Lorentz}(1896)}]{Lorentz:1896}%
  \BibitemOpen
  \bibfield  {author} {\bibinfo {author} {\bibfnamefont {H.~A.}\ \bibnamefont
  {Lorentz}},\ }\href@noop {} {\bibfield  {journal} {\bibinfo  {journal}
  {Amsterdammer Akademie der Wetenschappen}\ }\textbf {\bibinfo {volume} {4}},\
  \bibinfo {pages} {176} (\bibinfo {year} {1896})}\BibitemShut {NoStop}%
\bibitem [{\citenamefont {Buhmann}\ \emph {et~al.}(2012)\citenamefont
  {Buhmann}, \citenamefont {Butcher},\ and\ \citenamefont
  {Scheel}}]{Buhmann_nonreciprocal:2012}%
  \BibitemOpen
  \bibfield  {author} {\bibinfo {author} {\bibfnamefont {S.~Y.}\ \bibnamefont
  {Buhmann}}, \bibinfo {author} {\bibfnamefont {D.~T.}\ \bibnamefont
  {Butcher}}, \ and\ \bibinfo {author} {\bibfnamefont {S.}~\bibnamefont
  {Scheel}},\ }\href
  {http://iopscience.iop.org/article/10.1088/1367-2630/14/8/083034/pdf}
  {\bibfield  {journal} {\bibinfo  {journal} {New Journal of Physics}\ }\textbf
  {\bibinfo {volume} {14}},\ \bibinfo {pages} {083034} (\bibinfo {year}
  {2012})}\BibitemShut {NoStop}%
\bibitem [{\citenamefont {Yannopapas}(2006)}]{Yannopapas:2006}%
  \BibitemOpen
  \bibfield  {author} {\bibinfo {author} {\bibfnamefont {V.}~\bibnamefont
  {Yannopapas}},\ }\href
  {http://iopscience.iop.org/article/10.1088/0953-8984/18/29/025/meta;jsessionid=F1D4F70682DB34BAA10BD5D6CF1BA164.c3.iopscience.cld.iop.org}
  {\bibfield  {journal} {\bibinfo  {journal} {Journal of Physics: Condensed
  Matter}\ }\textbf {\bibinfo {volume} {18}},\ \bibinfo {pages} {6883}
  (\bibinfo {year} {2006})}\BibitemShut {NoStop}%
\bibitem [{\citenamefont {Thiel}\ \emph {et~al.}(2009)\citenamefont {Thiel},
  \citenamefont {Rill}, \citenamefont {von Freymann},\ and\ \citenamefont
  {Wegener}}]{Thiel:2009}%
  \BibitemOpen
  \bibfield  {author} {\bibinfo {author} {\bibfnamefont {M.}~\bibnamefont
  {Thiel}}, \bibinfo {author} {\bibfnamefont {M.~S.}\ \bibnamefont {Rill}},
  \bibinfo {author} {\bibfnamefont {G.}~\bibnamefont {von Freymann}}, \ and\
  \bibinfo {author} {\bibfnamefont {M.}~\bibnamefont {Wegener}},\ }\href
  {https://www.researchgate.net/profile/Georg_Von_Freymann/publication/229464911_ThreeDimensional_BiChiral_Photonic_Crystals/links/00b49537f7606b227b000000.pdf}
  {\bibfield  {journal} {\bibinfo  {journal} {Advanced Materials}\ }\textbf
  {\bibinfo {volume} {21}},\ \bibinfo {pages} {4680} (\bibinfo {year}
  {2009})}\BibitemShut {NoStop}%
\bibitem [{\citenamefont {Shen}(2014)}]{Shen:2014}%
  \BibitemOpen
  \bibfield  {author} {\bibinfo {author} {\bibfnamefont {S.-Q.}\ \bibnamefont
  {Shen}},\ }\href@noop {} {\bibfield  {journal} {\bibinfo  {journal} {National
  Science Review}\ }\textbf {\bibinfo {volume} {1}},\ \bibinfo {pages} {49}
  (\bibinfo {year} {2014})}\BibitemShut {NoStop}%
\bibitem [{\citenamefont {Bernevig}\ \emph {et~al.}(2006)\citenamefont
  {Bernevig}, \citenamefont {Hughes},\ and\ \citenamefont
  {Zhang}}]{Bernevig:2006}%
  \BibitemOpen
  \bibfield  {author} {\bibinfo {author} {\bibfnamefont {B.~A.}\ \bibnamefont
  {Bernevig}}, \bibinfo {author} {\bibfnamefont {T.~L.}\ \bibnamefont
  {Hughes}}, \ and\ \bibinfo {author} {\bibfnamefont {S.~C.}\ \bibnamefont
  {Zhang}},\ }\href {http://arxiv.org/pdf/cond-mat/0611399.pdf} {\bibfield
  {journal} {\bibinfo  {journal} {Science}\ }\textbf {\bibinfo {volume}
  {314}},\ \bibinfo {pages} {1757} (\bibinfo {year} {2006})}\BibitemShut
  {NoStop}%
\bibitem [{\citenamefont {Hasan}\ and\ \citenamefont
  {Kane}(2010)}]{Hasan:2010}%
  \BibitemOpen
  \bibfield  {author} {\bibinfo {author} {\bibfnamefont {M.~Z.}\ \bibnamefont
  {Hasan}}\ and\ \bibinfo {author} {\bibfnamefont {C.~L.}\ \bibnamefont
  {Kane}},\ }\href {http://journals.aps.org/rmp/pdf/10.1103/RevModPhys.82.3045}
  {\bibfield  {journal} {\bibinfo  {journal} {Reviews of Modern Physics}\
  }\textbf {\bibinfo {volume} {82}},\ \bibinfo {pages} {3045} (\bibinfo {year}
  {2010})}\BibitemShut {NoStop}%
\bibitem [{\citenamefont {Zhang}\ \emph {et~al.}(2009)\citenamefont {Zhang},
  \citenamefont {X.}, \citenamefont {Qi}, \citenamefont {Dai}, \citenamefont
  {Fang},\ and\ \citenamefont {Zhang}}]{Zhang:2009}%
  \BibitemOpen
  \bibfield  {author} {\bibinfo {author} {\bibfnamefont {H.}~\bibnamefont
  {Zhang}}, \bibinfo {author} {\bibfnamefont {L.~C.}\ \bibnamefont {X.}},
  \bibinfo {author} {\bibfnamefont {X.~L.}\ \bibnamefont {Qi}}, \bibinfo
  {author} {\bibfnamefont {X.}~\bibnamefont {Dai}}, \bibinfo {author}
  {\bibfnamefont {Z.}~\bibnamefont {Fang}}, \ and\ \bibinfo {author}
  {\bibfnamefont {S.~C.}\ \bibnamefont {Zhang}},\ }\href@noop {} {\bibfield
  {journal} {\bibinfo  {journal} {Nature Physics}\ }\textbf {\bibinfo {volume}
  {5}},\ \bibinfo {pages} {438} (\bibinfo {year} {2009})}\BibitemShut {NoStop}%
\bibitem [{\citenamefont {Qi}\ \emph {et~al.}(2009)\citenamefont {Qi},
  \citenamefont {Li}, \citenamefont {Zang},\ and\ \citenamefont
  {Zhang}}]{Qi:2009}%
  \BibitemOpen
  \bibfield  {author} {\bibinfo {author} {\bibfnamefont {X.-L.}\ \bibnamefont
  {Qi}}, \bibinfo {author} {\bibfnamefont {R.}~\bibnamefont {Li}}, \bibinfo
  {author} {\bibfnamefont {J.}~\bibnamefont {Zang}}, \ and\ \bibinfo {author}
  {\bibfnamefont {S.-C.}\ \bibnamefont {Zhang}},\ }\href
  {https://www.researchgate.net/profile/Jiadong_Zang/publication/23959588_Inducing_a_magnetic_monopole_with_topological_surface_States/links/53ea28920cf2dc24b3cb0f85.pdf}
  {\bibfield  {journal} {\bibinfo  {journal} {Science}\ }\textbf {\bibinfo
  {volume} {323}},\ \bibinfo {pages} {1184} (\bibinfo {year}
  {2009})}\BibitemShut {NoStop}%
\bibitem [{\citenamefont {Chen}\ \emph {et~al.}(2010)\citenamefont {Chen},
  \citenamefont {Chu}, \citenamefont {Analytis}, \citenamefont {Liu},
  \citenamefont {Igarashi}, \citenamefont {Kuo}, \citenamefont {Qi},
  \citenamefont {Mo}, \citenamefont {G.}, \citenamefont {Lu}, \citenamefont
  {Hashimoto}, \citenamefont {Sasagawa}, \citenamefont {Zhang}, \citenamefont
  {Fisher}, \citenamefont {Hussain},\ and\ \citenamefont {Shen}}]{Chen:2010}%
  \BibitemOpen
  \bibfield  {author} {\bibinfo {author} {\bibfnamefont {Y.~L.}\ \bibnamefont
  {Chen}}, \bibinfo {author} {\bibfnamefont {J.-H.}\ \bibnamefont {Chu}},
  \bibinfo {author} {\bibfnamefont {J.~G.}\ \bibnamefont {Analytis}}, \bibinfo
  {author} {\bibfnamefont {Z.~K.}\ \bibnamefont {Liu}}, \bibinfo {author}
  {\bibfnamefont {K.}~\bibnamefont {Igarashi}}, \bibinfo {author}
  {\bibfnamefont {H.-H.}\ \bibnamefont {Kuo}}, \bibinfo {author} {\bibfnamefont
  {X.~L.}\ \bibnamefont {Qi}}, \bibinfo {author} {\bibfnamefont {S.~K.}\
  \bibnamefont {Mo}}, \bibinfo {author} {\bibfnamefont {M.~R.}\ \bibnamefont
  {G.}}, \bibinfo {author} {\bibfnamefont {D.~H.}\ \bibnamefont {Lu}}, \bibinfo
  {author} {\bibfnamefont {M.}~\bibnamefont {Hashimoto}}, \bibinfo {author}
  {\bibfnamefont {T.}~\bibnamefont {Sasagawa}}, \bibinfo {author}
  {\bibfnamefont {S.~C.}\ \bibnamefont {Zhang}}, \bibinfo {author}
  {\bibfnamefont {I.~R.}\ \bibnamefont {Fisher}}, \bibinfo {author}
  {\bibfnamefont {Z.}~\bibnamefont {Hussain}}, \ and\ \bibinfo {author}
  {\bibfnamefont {Z.~X.}\ \bibnamefont {Shen}},\ }\href
  {http://science.sciencemag.org/content/329/5992/659.article-info} {\bibfield
  {journal} {\bibinfo  {journal} {Science}\ }\textbf {\bibinfo {volume}
  {329}},\ \bibinfo {pages} {659} (\bibinfo {year} {2010})}\BibitemShut
  {NoStop}%
\bibitem [{\citenamefont {Maciejko}\ \emph {et~al.}(2010)\citenamefont
  {Maciejko}, \citenamefont {Qi}, \citenamefont {Drew},\ and\ \citenamefont
  {Zhang}}]{Maciejko:2010}%
  \BibitemOpen
  \bibfield  {author} {\bibinfo {author} {\bibfnamefont {J.}~\bibnamefont
  {Maciejko}}, \bibinfo {author} {\bibfnamefont {X.-L.}\ \bibnamefont {Qi}},
  \bibinfo {author} {\bibfnamefont {H.~D.}\ \bibnamefont {Drew}}, \ and\
  \bibinfo {author} {\bibfnamefont {S.-C.}\ \bibnamefont {Zhang}},\ }\href
  {http://journals.aps.org/prl/pdf/10.1103/PhysRevLett.105.166803} {\bibfield
  {journal} {\bibinfo  {journal} {Physical Review Letters}\ }\textbf {\bibinfo
  {volume} {105}},\ \bibinfo {pages} {166803} (\bibinfo {year}
  {2010})}\BibitemShut {NoStop}%
\bibitem [{\citenamefont {Wilczek}(1987)}]{Wilczek:1987}%
  \BibitemOpen
  \bibfield  {author} {\bibinfo {author} {\bibfnamefont {F.}~\bibnamefont
  {Wilczek}},\ }\href
  {http://journals.aps.org/prl/pdf/10.1103/PhysRevLett.58.1799} {\bibfield
  {journal} {\bibinfo  {journal} {Physical Review Letters}\ }\textbf {\bibinfo
  {volume} {58}},\ \bibinfo {pages} {1799} (\bibinfo {year}
  {1987})}\BibitemShut {NoStop}%
\bibitem [{\citenamefont {Grushin}\ and\ \citenamefont
  {Cortijo}(2011)}]{Grushin:2011}%
  \BibitemOpen
  \bibfield  {author} {\bibinfo {author} {\bibfnamefont {A.~G.}\ \bibnamefont
  {Grushin}}\ and\ \bibinfo {author} {\bibfnamefont {A.}~\bibnamefont
  {Cortijo}},\ }\href
  {http://journals.aps.org/prl/pdf/10.1103/PhysRevLett.106.020403} {\bibfield
  {journal} {\bibinfo  {journal} {Physical Review Letters}\ }\textbf {\bibinfo
  {volume} {106}},\ \bibinfo {pages} {020403} (\bibinfo {year}
  {2011})}\BibitemShut {NoStop}%
\bibitem [{\citenamefont {Chang}\ and\ \citenamefont
  {Yang}(2009)}]{Chang:2009}%
  \BibitemOpen
  \bibfield  {author} {\bibinfo {author} {\bibfnamefont {M.-C.}\ \bibnamefont
  {Chang}}\ and\ \bibinfo {author} {\bibfnamefont {M.-F.}\ \bibnamefont
  {Yang}},\ }\href {http://journals.aps.org/prb/pdf/10.1103/PhysRevB.80.113304}
  {\bibfield  {journal} {\bibinfo  {journal} {Physical Review B}\ }\textbf
  {\bibinfo {volume} {80}},\ \bibinfo {pages} {113304} (\bibinfo {year}
  {2009})}\BibitemShut {NoStop}%
\bibitem [{\citenamefont {Zuo}\ \emph {et~al.}(2013)\citenamefont {Zuo},
  \citenamefont {Ling}, \citenamefont {Sheng},\ and\ \citenamefont
  {Xing}}]{Zuo:2013}%
  \BibitemOpen
  \bibfield  {author} {\bibinfo {author} {\bibfnamefont {Z.~W.}\ \bibnamefont
  {Zuo}}, \bibinfo {author} {\bibfnamefont {D.~B.}\ \bibnamefont {Ling}},
  \bibinfo {author} {\bibfnamefont {L.}~\bibnamefont {Sheng}}, \ and\ \bibinfo
  {author} {\bibfnamefont {D.~Y.}\ \bibnamefont {Xing}},\ }\href@noop {}
  {\bibfield  {journal} {\bibinfo  {journal} {Physics Letters A}\ }\textbf
  {\bibinfo {volume} {377}},\ \bibinfo {pages} {2909} (\bibinfo {year}
  {2013})}\BibitemShut {NoStop}%
\bibitem [{\citenamefont {Crosse}(2015)}]{Crosse_2:2015}%
  \BibitemOpen
  \bibfield  {author} {\bibinfo {author} {\bibfnamefont {J.~A.}\ \bibnamefont
  {Crosse}},\ }\href {http://arxiv.org/ftp/arxiv/papers/1510/1510.06130.pdf}
  {\bibfield  {journal} {\bibinfo  {journal} {arXiv preprint arXiv:1510.06130}\
  } (\bibinfo {year} {2015})}\BibitemShut {NoStop}%
\bibitem [{\citenamefont {Rosa}\ \emph
  {et~al.}(2008{\natexlab{a}})\citenamefont {Rosa}, \citenamefont {Dalvit},\
  and\ \citenamefont {Milonni}}]{Rosa:2008}%
  \BibitemOpen
  \bibfield  {author} {\bibinfo {author} {\bibfnamefont {F.~S.~S.}\
  \bibnamefont {Rosa}}, \bibinfo {author} {\bibfnamefont {D.~A.~R.}\
  \bibnamefont {Dalvit}}, \ and\ \bibinfo {author} {\bibfnamefont {P.~W.}\
  \bibnamefont {Milonni}},\ }\href
  {http://journals.aps.org/prl/pdf/10.1103/PhysRevLett.100.183602} {\bibfield
  {journal} {\bibinfo  {journal} {Physical Review Letters}\ }\textbf {\bibinfo
  {volume} {100}},\ \bibinfo {pages} {183602} (\bibinfo {year}
  {2008}{\natexlab{a}})}\BibitemShut {NoStop}%
\bibitem [{\citenamefont {Rosa}\ \emph
  {et~al.}(2008{\natexlab{b}})\citenamefont {Rosa}, \citenamefont {Dalvit},\
  and\ \citenamefont {Milonni}}]{Rosa:2008_2}%
  \BibitemOpen
  \bibfield  {author} {\bibinfo {author} {\bibfnamefont {F.~S.~S.}\
  \bibnamefont {Rosa}}, \bibinfo {author} {\bibfnamefont {D.~A.~R.}\
  \bibnamefont {Dalvit}}, \ and\ \bibinfo {author} {\bibfnamefont {P.~W.}\
  \bibnamefont {Milonni}},\ }\href
  {http://journals.aps.org/pra/pdf/10.1103/PhysRevA.78.032117} {\bibfield
  {journal} {\bibinfo  {journal} {Physical Review A}\ }\textbf {\bibinfo
  {volume} {78}},\ \bibinfo {pages} {032117} (\bibinfo {year}
  {2008}{\natexlab{b}})}\BibitemShut {NoStop}%
\bibitem [{\citenamefont {Zhao}\ \emph {et~al.}(2009)\citenamefont {Zhao},
  \citenamefont {Zhou}, \citenamefont {Koschny}, \citenamefont {Economou},\
  and\ \citenamefont {Soukoulis}}]{Zhao:2009}%
  \BibitemOpen
  \bibfield  {author} {\bibinfo {author} {\bibfnamefont {R.}~\bibnamefont
  {Zhao}}, \bibinfo {author} {\bibfnamefont {J.}~\bibnamefont {Zhou}}, \bibinfo
  {author} {\bibfnamefont {T.}~\bibnamefont {Koschny}}, \bibinfo {author}
  {\bibfnamefont {E.~N.}\ \bibnamefont {Economou}}, \ and\ \bibinfo {author}
  {\bibfnamefont {C.~M.}\ \bibnamefont {Soukoulis}},\ }\href
  {http://journals.aps.org/prl/pdf/10.1103/PhysRevLett.103.103602} {\bibfield
  {journal} {\bibinfo  {journal} {Physical Review Letters}\ }\textbf {\bibinfo
  {volume} {103}},\ \bibinfo {pages} {103602} (\bibinfo {year}
  {2009})}\BibitemShut {NoStop}%
\bibitem [{\citenamefont {Chen}\ and\ \citenamefont {Wan}(2011)}]{Chen:2011}%
  \BibitemOpen
  \bibfield  {author} {\bibinfo {author} {\bibfnamefont {L.}~\bibnamefont
  {Chen}}\ and\ \bibinfo {author} {\bibfnamefont {S.}~\bibnamefont {Wan}},\
  }\href {http://link.aps.org/doi/10.1103/PhysRevB.84.075149} {\bibfield
  {journal} {\bibinfo  {journal} {Phys. Rev. B}\ }\textbf {\bibinfo {volume}
  {84}},\ \bibinfo {pages} {075149} (\bibinfo {year} {2011})}\BibitemShut
  {NoStop}%
\bibitem [{\citenamefont {Cysne}\ \emph {et~al.}(2014)\citenamefont {Cysne},
  \citenamefont {Kort-Kamp}, \citenamefont {Oliver}, \citenamefont {Pinheiro},
  \citenamefont {Rosa},\ and\ \citenamefont {Farina}}]{Cysne:2014}%
  \BibitemOpen
  \bibfield  {author} {\bibinfo {author} {\bibfnamefont {T.}~\bibnamefont
  {Cysne}}, \bibinfo {author} {\bibfnamefont {W.~J.~M.}\ \bibnamefont
  {Kort-Kamp}}, \bibinfo {author} {\bibfnamefont {D.}~\bibnamefont {Oliver}},
  \bibinfo {author} {\bibfnamefont {F.~A.}\ \bibnamefont {Pinheiro}}, \bibinfo
  {author} {\bibfnamefont {F.~S.~S.}\ \bibnamefont {Rosa}}, \ and\ \bibinfo
  {author} {\bibfnamefont {C.}~\bibnamefont {Farina}},\ }\href@noop {}
  {\bibfield  {journal} {\bibinfo  {journal} {Phys. Rev. A}\ }\textbf {\bibinfo
  {volume} {90}},\ \bibinfo {pages} {052511} (\bibinfo {year}
  {2014})}\BibitemShut {NoStop}%
\bibitem [{\citenamefont {Crosse}\ \emph {et~al.}(2015)\citenamefont {Crosse},
  \citenamefont {Fuchs},\ and\ \citenamefont {Buhmann}}]{Crosse:2015}%
  \BibitemOpen
  \bibfield  {author} {\bibinfo {author} {\bibfnamefont {J.~A.}\ \bibnamefont
  {Crosse}}, \bibinfo {author} {\bibfnamefont {S.}~\bibnamefont {Fuchs}}, \
  and\ \bibinfo {author} {\bibfnamefont {S.~Y.}\ \bibnamefont {Buhmann}},\
  }\href {\doibase 10.1103/PhysRevA.92.063831} {\bibfield  {journal} {\bibinfo
  {journal} {Physical Review A}\ }\textbf {\bibinfo {volume} {92}},\ \bibinfo
  {pages} {063831} (\bibinfo {year} {2015})}\BibitemShut {NoStop}%
\bibitem [{\citenamefont {Curie}(1894)}]{Curie:1894}%
  \BibitemOpen
  \bibfield  {author} {\bibinfo {author} {\bibfnamefont {P.}~\bibnamefont
  {Curie}},\ }\href@noop {} {\bibfield  {journal} {\bibinfo  {journal} {J.
  Phys. Theor. Appl.}\ }\textbf {\bibinfo {volume} {3}},\ \bibinfo {pages}
  {393} (\bibinfo {year} {1894})}\BibitemShut {NoStop}%
\bibitem [{\citenamefont {Barton}(1970)}]{Barton:1970}%
  \BibitemOpen
  \bibfield  {author} {\bibinfo {author} {\bibfnamefont {G.}~\bibnamefont
  {Barton}},\ }\href@noop {} {\bibfield  {journal} {\bibinfo  {journal} {Proc.
  Roy. Soc. Lond. A.}\ }\textbf {\bibinfo {volume} {320}},\ \bibinfo {pages}
  {251} (\bibinfo {year} {1970})}\BibitemShut {NoStop}%
\bibitem [{\citenamefont {Barton}(1979)}]{Barton:1979}%
  \BibitemOpen
  \bibfield  {author} {\bibinfo {author} {\bibfnamefont {G.}~\bibnamefont
  {Barton}},\ }\href@noop {} {\bibfield  {journal} {\bibinfo  {journal} {Proc.
  R. Soc. Lond. A.}\ }\textbf {\bibinfo {volume} {367}},\ \bibinfo {pages}
  {117} (\bibinfo {year} {1979})}\BibitemShut {NoStop}%
\bibitem [{\citenamefont {Bates}\ and\ \citenamefont
  {Bederson}(1991)}]{Hinds_Book}%
  \BibitemOpen
  \bibinfo {editor} {\bibfnamefont {D.}~\bibnamefont {Bates}}\ and\ \bibinfo
  {editor} {\bibfnamefont {B.}~\bibnamefont {Bederson}},\ eds.,\ \href@noop {}
  {\emph {\bibinfo {title} {Advances in Atomic, Molecular, and Optical
  Physics}}},\ Vol.~\bibinfo {volume} {28}\ (\bibinfo  {publisher} {Academic
  Press, INC.},\ \bibinfo {year} {1991})\BibitemShut {NoStop}%
\end{thebibliography}

%

\end{document}